\documentstyle[12pt,epsf]{article}

\input{psfig}

\newcommand{\nc}{\newcommand}
\def\frac#1#2{{\textstyle {#1 \over #2}}}

\nc{\beq}{\begin{equation}}
\nc{\eeq}{\end{equation}}
\nc{\beqa}{\begin{eqnarray}}
\nc{\eeqa}{\end{eqnarray}}
\nc{\lsim}{\begin{array}{c}\,\sim\vspace{-21pt}\\< \end{array}}
\nc{\gsim}{\begin{array}{c}\sim\vspace{-21pt}\\> \end{array}}
\nc{\eps}{\epsilon}
\nc{\s}{\sigma}
\nc{\veps}{\varepsilon}
\nc{\no}{\noindent}
\nc{\D}{\Delta}
\nc{\nn}{\nonumber}
\nc{\al}{\alpha}
\nc{\be}{\beta}
\nc{\ga}{\gamma}
\nc{\de}{\delta}
\nc{\w}{\wedge}
\nc{\la}{\lambda}
\nc{\om}{\omega}
\nc{\Om}{\Omega}
\nc{\p}{\partial}
\nc{\lra}{\longrightarrow}
\nc{\Lra}{\Longrightarrow}
\nc{\equi}{\Longleftrightarrow}
\nc{\dmi}{{1\over 2}}

\begin{document}

\begin{center}
\vskip .5 in
{\large \bf 
A Semi-Classical Analysis of Order from Disorder}

\vskip .6 in

  {\bf B. Dou\c{c}ot}\footnote{doucot@lpthe.jussieu.fr} and  {\bf P. Simon}\footnote{simon@lpthe.jussieu.fr }
   \vskip 0.3 cm
 {\it   Laboratoire de Physique Th\'eorique et Hautes Energies  }\footnote{ 
 Unit\'e associ\'ee au CNRS URA 280}\\
 {\it  Universit\'es Pierre et Marie Curie Paris VI et Denis Diderot Paris 
VII}\\
{\it  2 pl. Jussieu, 75251 Paris cedex 05 }

  \vskip  1cm   
\end{center}
\vskip .5 in

 \begin{abstract}
We study in this paper the Heisenberg antiferromagnet with nearest neighbours interactions on the Husimi cactus, a system which has locally the same topology as the Kagom\'e lattice.
This system has a huge classical  degeneracy corresponding to an extensive number of degrees of freedom.
We show that unlike  thermal fluctuations, quantum fluctuations lift partially this degeneracy and favour a discrete subset of classical ground states. In order to clarify the origin of these effects, we have set up a general semi-classical analysis of the order from disorder phenomenon and clearly identified the differences between classical and quantum fluctuations. This semi-classical approach also enables us to classify various situations where a selection mechanism still occurs.
Moreover, once a discrete set of ground states has been preselected, our analysis suggests that tunelling processes within this set should be the dominant effect underlying  the strange low energy spectrum of  Kagom\'e-like  lattices. 

 \end{abstract}

\no 

\eject
\renewcommand{\thepage}{\arabic{page}}
\setcounter{page}{1}
\setcounter{equation}{0}
\noindent

\section{Introduction}

Motivated by the search and understanding of new quantum disordered ground states in magnetic systems, recent years have seen a renewal of interest on the properties of 
 frustrated Heisenberg antiferromagnetic (HAF) systems with a special attention to their possible relations with superconductivity. More specifically, a considerable amount of work has been done on frustrated systems whose classical ground states exhibit an infinite number of local continuous degeneracies. Typical examples of such systems are the HAF models on the Kagom\'e lattice in two dimensions ($2D$) and the HAF models on the pyrochlore lattice in three dimensions. The HAF models on the Kagom\'e lattice has been originally inspired by the experiments on the $^3$He layer on graphite \cite{greywall,franco} and also by the structure of the $SC_rGO$ compound \cite{ramirez}.
Other compounds  whose magnetic ions live on a pyrochlore lattice have also been experimentaly studied by neutron scattering \cite{gaulin}.
From the theoretical point of view, the ground state of  the quantum HAF model on the Kagom\'e lattice  is believed to be a quantum spin liquid. This model has been investigated by exact diagonalizations on finite lattices \cite{diag}, series expansions
\cite{serie}, Large-$N$ calculations \cite{sachdev} and also with semi-classical approaches
\cite{harris92,chubukov92,chalker92,chandra93}. More recently, some theoretical works on the HAF model on the pyrochlore lattice have pointed out that this model is already disordered at any finite temperature \cite{chalker97} unlike the xlassical Kagom\'e Heisenberg antiferromagnet which chooses coplanar spin configurations at low temperatures.
The low energy spectrum of the HAF model  with spin $\dmi$ on the Kagom\'e lattice \cite{lechem} and the pyrochlore lattice \cite{canals} have been analyzed by performing exact diagonalizations on small clusters. The results are quite fascinating in both cases since there is a singlet-triplet gap and nevertheless a large singlet degeneracy which increases with the cluster size. It seems then plausible that the huge classical degeneracy should play a major role in order to explain these surprising results.

Systems with many degenerate classical ground-states
were often considered as useful toy models before studying for instance
spin glasses for which {\em both} frustration and randomness are relevant.
It has been then noticed by J. Villain and coworkers \cite{vil1}, 
and also by E. Shender \cite{order},
that switching on any source of fluctuation (typically, either classical
thermal or quantum zero-point fluctuations) has a tendancy to lift the degeneracy
between these classical ground-states. Often this mechanism picks ordered
ground-states since they exhibit more symmetries and therefore encourage
larger fluctuations. This is the so-called ``order from disorder'' phenomenon.
Over the last ten years, many interesting contributions have been dedicated to
this topic, in the context of the frustrated square lattice \cite{henley},
the Kagom\'{e} lattice \cite{chalker92,chandra93}, and the pyrochlore lattice \cite{chalker97,canals}.
A common trend in all these works is the appearance of at least a partial 
degeneracy lifting within the ground-state manifold, in the presence of classical
thermal or quantum zero-point fluctuations. In most cases, both types of fluctuations
favour the same subset of ground-states. This appears to be a little surprising,
since we shall provide in this paper an explicit example for which they behave in
a qualitatively different way. But the question of degeneracy lifting is only
one part of the story, since it is mostly a {\em local} analysis in the
vicinity of a given ground-state. Maybe more crucial to the physical behaviour
is the {\em global} question of whether the system remains close to any of the 
favoured ground-states or not. This question is closely connected to the size
of energy barriers separating these pre-selected ground-states. If these barriers
are too small, either thermal activation or quantum tunneling may delocalize the
system in phase-space, and thus destroy the selection itself. In fact, evidence
now strongly accumulates to support such a scenario, either for the Kagom\'e
or the pyrochlore lattices, which exhibit classical and quantum spin liquid phases
with apparently no long-range order. Although this global aspect of the
selection mechanism is very interesting, there are few systems for which a
rather complete analytical treatment is available. Therefore, in
spite of the fact that different systems often exhibit different behaviours,
we believe that valuable insight may be gained from simpler models which can be
studied more easily than more realistic geometries. 

Among these, the Husimi cactus
(see Figures 3 and 4) has already provided an explicit example of a system with a
continuous manifold of classical ground-states where energy barriers are too low to
freeze the system in the vicinity of a given ground-state in the classical thermal
case, or to prevent dramatic tunneling processes in the $T=0$ case \cite{husim}.
The relevance of tunneling is clearly demonstrated by comparing the relatively
small ground-state degeneracy (for spin $S=1/2$) which scales as the number of sites
and the huge classical degeneracy: the number of coplanar spin ground-states is
for instance growing exponentially with the number of sites. The question we address
here is: tunneling between {\em which} classical ground-states? More precisely:
does the quantum zero point energy depend on the classical ground-state? We have found
that unlike classical thermal fluctuations for which all classical ground-states
on the Husimi cactus are equivalent, quantum fluctuations do favour the discrete
subset of coplanar classical ground-states. We believe it is one of the clearest
examples where thermal and quantum fluctuations behave in a qualitatively different
way, already at the non-interacting spin-wave level. In order to clarify the origin of
this effect, we have set-up a semi-classical analysis of order from disorder.
The main idea is simply to calculate the spin-wave spectrum from the classical equations
of motion for spins, since the classical frequency also determines the quantum zero-point motion energy of a harmonic mode. The classical dynamical approach greatly simplifies
for instance the separation between fast oscillations around a classical ground-state and
slow drift motions along the classical ground-state manifold, which are connected to zero
modes. It also provides a very simple understanding for the different nature of classical
thermal and quantum zero point fluctuations. In the context of the Husimi cactus, the
equivalence between all the ground-states for thermal fluctuations is related to the
presence of continuous transformations which conserve the total energy and phase-space
volume, and therefore the Gibbs measure. An example of such a transformation is a rotation of 
all the spins below a given site on the cactus around the spin at this given site.
However, conservation of the Gibbs measure is not sufficient to enforce conservation of
the zero-point motion energy, since hamiltonian dynamics requires also a symplectic structure
on phase-space (Poisson brackets), which turns out not to be preserved by the aforementioned
transformations. We shall show that the non-conservation of Poisson brackets under these
phase-space transformations is due to the fact that they cannot be generated (via Poisson
brackets) by functions defined on phase-space. This shows that it is rather difficult to
preserve quantum-mechanically a continuous classical degeneracy which is not enforced by
a symmetry of the Hamiltonian. We expect that only a discrete residual degeneracy survives
after introducing quantum zero point motion. 

We believe this analysis contributes to rule out one possible scenario 
which might have been advocated to understand the unusual situation
of the spin $1/2$ Kagom\'e lattice. The generation of very low energy scales
in the singlet sector might have been attributed to very slow drift motions
close to and along the continuous set of classical ground-states. Note that this
continuous set is {\em not} a smooth manifold for either
the Kagom\'e or the pyrochlore lattices, since for those the
number of zero modes depends on the classical ground-state \cite{chalker92}. Although
simple and appealing, this idea does not work for two reasons.
The first one is the large gap between the singlet and the spin one sectors,
which shows the impossibility to construct long-lived wave packets centered mostly
around a single classical ground-state. The other reason is provided by our analysis:
because of quantum zero-point fluctuations, the ground-state continuum turns into 
a discrete set, so that classical drift motions in the vicinity of this continuum
acquire finite typical frequencies. Therefore, we conclude that a semi-classical
analysis of the energy spectrum on Kagom\'e-like lattices  (of which the Husimi cactus 
is a simple example) should focus on tunneling processes within a discrete set of preselected
(via quantum fluctuations) classical ground-states. These processes are crucial to
induce the gap in the spin one sector. The main challenge for the Kagom\'e lattice is then
to understand why they  lift so weakly the degeneracy in the singlet sector.
We haven't addressed this crucial question here, but we have clarified  what happens
when we try to quantize a system with a smooth manifold of classical ground-states,
whose dimension is larger than the number of independent symmetries of the Hamiltonian.
Either tunneling plays a minor role, but then the classical slow variables experience an effective 
energy landscape which is no longer flat, or tunneling is the dominant effect, as we believe
is the case for both the Husimi cactus and the Kagom\'e lattice.

Our presentation is organized as follows. Section 2 will present very simple examples with
two degrees of freedom and a one dimensional degenerate ground-state manifold. This 
will illustrate how order form disorder  works in a classical hamiltonian system,
provided the fast modes are prepared with a finite value of the corresponding action variable.
Section 3 is dedicated to a simple spin chain for special values of coupling
parameters which produce a  more symmetrical ground-state manifold (the full sphere $S^{2}$) than
one would expect from the symmetries of the Hamiltonian (rotations around a single axis).
Section 4 presents the main results for the Husimi cactus, and section 5 contains
our geometrical interpretation of the difference between thermal classical and  quantum zero-point
fluctuations. A brief conclusion is given in section 6, and several appendices present
some technical details that most readers will prefer to skip.

\section{Some simple examples with two degrees of freedom}
\setcounter{equation}{0}

\noindent
In this section, we consider simple examples of dynamical systems where the classical
ground-state manifold is one dimensional. This will demonstrate explicitely that classical
dynamics exhibits the phenomenon of order from disorder provided the fast modes are
``fast enough''.

\subsection{An integrable toy model}

\noindent
Let us consider a dynamical system with two degrees of freedom, with coordinates
$(X_{1},P_{1},X_{2},P{_2})$ on phase space. Here $(X_{1},P_{1})$ denote a slow
mode which is continuously degenerate along the line $(X_{1},0,0,0)$ in phase space,
and $(X_{2},P_{2})$ a fast mode, which oscillation frequency $\Om$ is a smooth function 
of $X_{1}$. We thus introduce the following Hamiltonian:
\beq
H=\frac{1}{2} P_{1}^{2}+\frac{1}{2} \Om (X_{1}) (P_{2}^{2}+X_{2}^{2})
\eeq
We note that $P_{1}$ commutes with $H$ only if $\Om$ is a constant. So the classical degeneracy
is not in general due to a symmetry of $H$. This Hamiltonian is integrable since we
can perform the following canonical transformation $(X_{1},P_{1},X_{2},P{_2}) \lra 
(X_{1},P_{1},J_{2},\theta _{2})$, where $X_{2}= \sqrt{2J_{2}} \sin \theta _{2}$ and
$P_{2}= \sqrt{2J_{2}} \cos \theta _{2}$. With these new canonical variables, $H$ becomes:
\beq
H= \frac{1}{2} P_{1}^{2} + \Om (X_{1}) J_{2}
\eeq
The motion of the fast mode is described by:
\begin{eqnarray}
\dot{J}_{2} &=& 0 \\
\dot {\theta}_{2} &=& \Om (X_{1})
\end{eqnarray}
So, as we expect, the oscillation frequency is controlled by the ``slow'' variable
$X_{1}$, and is therefore time-dependent in general. Here the action $J_{2}$ is
{\em exactly} conserved, which is a special feature of this model.
In general, $J_{2}$ may be approximately conserved with good accuracy, since it is
an adiabatic invariant \cite{arnold1,landau}. However, this requires a good separation
between the slow and the fast time scales. This question will be addressed soon.
Turning now to the ``slow'' variables, we have:
\begin{eqnarray}
\dot{X}_{1} &=& P_{1} \\
\dot{P}_{1} &=& -J_{2} \frac{d\Om}{dX_{1}} (X_{1}) 
\end{eqnarray}
We see that if $J_{2}$ is non-vanishing, the slow variables $(X_{1},P_{1})$
acquire a non-trivial dynamics, with an effective potential given by:
$V_{eff}(X_{1})=J_{2} \Om (X_{1})$. Therefore the classical degeneracy is dynamically lifted,
and $X_{1}$ oscillates around the value which minimizes the fast frequency
$\Om (X_{1})$. In a semi-classical approach, the fast oscillator is excited with
quantized values for $J_{2}=\hbar (n+1/2)$, where $n$ is a non-negative 
integer.
So the effective potential corresponding to quantum zero-point motion is
obtained from $J_{2}= \frac{\hbar}{2}$. Although very simple, this model
captures the main idea we need throughout the rest of the paper.

\subsection{A more generic example}

\noindent
The special status of the previous model comes from the fact that
action-angle variables corresponding to the fast motion do not
depend on the slow variables $(X_{1},P_{1})$. In quantum mechanical words,
only the frequency of the fast motion depends on $X_{1}$, but the 
eigenstate basis of the corresponding harmonic oscillator doesn't.
In such a case, the adiabatic principle is exact, and eigenstates 
of the time-dependent problem are given by those of the time
independent one (up to phase factors). But in general, we expect 
the local eigenstate basis to depend on the actual value of $X_{1}$.
As a result, the classical action is only approximately conserved,
and therefore transitions to excited states may occur for the fast
oscillator, specially if the two frequency scales are not very well
separated. We show here a simple example where the adiabatic principle
may break down. We think this is a very interesting problem in classical
mechanics, which deserves further investigation. But this would be beyond 
the scope of this paper. We just provide here an illustration that
complicated things may happen. Let us then consider the following Hamiltonian:
\beq
H=\frac{1}{2} P_{1}^{2}+\frac{1}{2} P_{2}^{2}+ 
\frac{1}{2} (1+\epsilon X_{1}^{2}) X_{2}^{2}
\eeq
This represents as before a potential landscape similar to a gutter with
variable curvature depending on $X_{1}$. Assume first that $X_{1}$ can
be frozen. The fast oscillator evolves with a frequency
$\Om (X_{1}) = (1+\epsilon X_{1}^{2})^{1/2}$. The corresponding action-angle
variables $(J_{2}, \theta _{2})$ would now be given by:
\begin{eqnarray}
X_{2} &=& (\frac {2J_{2}}{\Om (X_{1})})^{1/2} \sin \theta _{2} \\
P_{2} &=& (2J_{2} \Om (X_{1}))^{1/2} \cos \theta _{2}
\end{eqnarray}
But now, since the definition of $(J_{2}, \theta _{2})$ involves explicitely
$X_{1}$, the corresponding transformation is no longer canonical.
As before $\frac {1}{2} P_{2}^{2} + \frac {1}{2} \Om ^{2}(X_{1}) X_{2}^{2}=
\Om (X_{1}) J_{2}$ is the effective potential of the slow variables, but 
$J_{2}$ is not exactly conserved. As usual, this effective Hamiltonian is obtained
by averaging over one period for the fast variables \cite{arnold1}. To evaluate
when the adiabatic approximation makes sense, let us compare the slow and the fast
frequencies. The averaged motion of the slow variables is given by:
\begin{eqnarray}
\dot{X}_{1} &=& P_{1} \\
\dot{P}_{1} &=& -J_{2} \frac{d\Om }{dX_{1}} (X_{1}) = 
-J_{2} \epsilon \frac {X_{1}}{(1+\epsilon X_{1}^{2})^(1/2)}
\end{eqnarray}
For the small oscillations of $X_{1}$, we find the slow frequency 
$\om = (J_{2} \epsilon)^{1/2}$ which has to be much smaller than the minimum
value of $\Om (X_{1})$, that is we require $J_{2} \epsilon \ll 1$. We have performed
a numerical integration of the equations of motion for this model in
order to check the overall picture. The results are summarized in Figures 1a,1b,1c,1d.
The Figures 1a and 1b represent the ``quasi orbits'' of the slow and fast oscillators for two different values of the parameter $\eps=0.1$ and $\eps=0.25$.
We clearly see a qualitative change in the dynamics as $\epsilon$ is increased
with a breakdown of the adiabatic approximation for larger values of $\epsilon$. This can be seen in a more explicit way by plotting directly the variations of the adiabatic invariant $J_2$ of the fast motion during one quasi period of the slow motion. This has been done in Figures 1c and 1d. The adiabatic approximation is no longer valid for $\eps=0.25$.
It would be very interesting to identify the underlying mechanism by which this
breakdown occurs. We just mention here that Figures  1c and 1d suggest that $J_{2}$ acquires
a time dependence with frequencies which are close to simple harmonics of the
``fast'' frequency. Therefore, a perturbative approach might give valuable
insights here.

 The data support however the validity of our picture up to rather
large values of $J_{2} \epsilon$, around $0.1$. We shall now move on to spin systems
with many degrees of freedom.

\section{A ferromagnetic chain}
\setcounter{equation}{0}

\subsection{Ground state and dispersion relations in the different phases}
\noindent
We study in this section the spin dynamics of a one dimensional XXZ chain 
with an anisotropic quadratic term. The Hamiltonian reads
\beq
\label{hamchain}
{\cal H}=\sum\limits_{i} J\left(S_i^xS_{i+1}^x+S_i^yS_{i+1}^y+\Delta S_i^zS_{i+1}^z\right)
+\sum\limits_i (S_i^z)^2
\eeq
where $\vec{S_i}$ is a three dimensional vector normalized to $S$ (the length of the vector is a constant of motion  from a classical and quantum point of view). 

The classical phase diagram of this model can be determined easily and has been drawn on 
Figure 2 (see also ref. \cite{denijs} for a quantum analysis with $S={1\over 2}$, $S=1$). 
It is composed of three different phases: the ferromagnetic one, the planar one 
and the antiferromagnetic one (AF).  In order to compute the low energy spectrum 
in the various  phases (at large $S$), it is easy to use the Holdstein-Primakoff transformation. 
Nevertheless, we prefer to work here with classical spin dynamics at $T=0$ in order 
to give a consistent treatment of the order from disorder phenomenon.  
This assumes a spin $\hat{S}$ reacts like a rotor which can be parametrized 
by a vector $\vec{S}$.
The evolution of a quantity $O(S)$ is given by:
$${dO(S)\over dt}= \{O,H\}$$ with $\{\cdots\}$  the Poisson brakets defined as: 
$$\{f,g\}=\eps^{abc} {\partial f\over \partial S^a}{\partial g\over \partial S^b}S^c$$
with $\eps^{abc}$ the completely antisymmetric tensor.
If $\vec{S}$ is a vector living on a sphere $S^2$, the corresponding classical Poisson brackets are given by: 
$\{S^a,S^b\}=\eps^{abc} S^c$. After a standard quantization process, we would obtain the usual commutation rules 
for spin operators (see \cite{spinq} for a rigorous treatment {\it \`a la Dirac}). The equations of motion then become:
\beq
\label{eqmvt}
{d S^a\over dt}=\eps^{abc} {\partial H\over \partial S^b} S^c
\eeq
These equations of motion can also be derived directly from a variational principle,
the latter being more adapted for a direct path integral quantization scheme.

We are now ready to derive the equations of motion following eq. (\ref{eqmvt}) for the hamiltonian (\ref{hamchain}) 
in each of the three possible phases.

In the ferromagnetic ground-state, we use the decomposition 
$\vec{S}=S^z \vec{z}+ \vec{S}^{\bot}$ (with \\$(\vec{z})^2=1$) and we obtain 
the system of equations:
\beqa
{d \vec{S}_n^{\bot}\over dt}&=&J\left[(\vec{S}_{n-1}^{\bot}+\vec{S}_{n+1}
^{\bot})\w S_n^z \vec{z}
+ \Delta( S_{n-1}^z+ S_{n+1}^z)\vec{z}\w \vec{S}_n\right.\nn\\
&&+\left. {D\over J}( S_{n}^z(\vec{z}\w \vec{S}_{n}^{\bot}))+(\vec{z}\w \vec{S}_{n}^{\bot})S_{n}^z)\right]\\
{d \vec{S}_n^z\over dt}&=&J\left(\vec{S}_{n-1}^{\bot}+\vec{S}_{n+1}^{\bot}\right)\w \vec{S}_n^{\bot}
\eeqa
In the vicinity of the ferromagnetic ground-state, we can assume:  
$$\vec{S}_n(t)=S\vec{z}+\exp{i(kn-\om t)}\vec{\veps}+O(\veps^2)$$
with $\vec{\veps}.\vec{z}=0$. By solving this system, we find the dispersion relation:
\beq
|\om (k)|=2JS|\Delta+{D\over J}-\cos(k)|
\eeq
We would have obtained the same low energy spectrum (with $\hbar=1$) by using for 
example the Holdstein-Primakoff transformation. When $\Delta+{D\over J}=-1$, 
the mode at $k=\pi$ becomes slow and has the
expected $k^{2}$ dispersion for a ferromagnetic ground-state.

In the planar region, we use the following general decomposition for $\vec{S}_n=\vec{S}_n^0+\delta\vec{S}_n$
\beq
\vec{S}_n=(-1)^nS\vec{u}+(A(-1)^n\vec{v}+B\vec{z})\exp{i(kn-\om t)}
\eeq
where $\vec{S}_n^0=(-1)^nS$. We have also defined
 $\vec{u}.\vec{z}=0,(\vec{u})^2=1$ and $\vec{v}=\vec{z}\w \vec{u}$. The equations giving $A$ and $B$ are:
\beqa
-i\om A &=& 2JS[1+\Delta\cos(ka)+{D\over J}]B\nn\\
i\om B &=& 2JS(1-\cos(ka))A
\eeqa
As $|\Delta|\le 1+{D\over J}$ in the planar region, we find:
\beq
\om =\pm2JS(1-\cos(ka))^{{1\over 2}}(1+\Delta\cos(ka)+{D\over J})^{{1\over 2}}
\eeq
Two interesting limits can be taken. First,
when $\Delta+{D\over J}\to -(1)^+$, we recover the soft mode in $k^2$, whereas  when  $\Delta-{D\over J}\to (1)^-$,  we find a linear dispersion around $k=0$ typical of an AF ground-state.

In the AF area, we can use the decomposition: 
\beq
\vec{S}_n=\vec{S}_n^z+\vec{S}_n^{\bot}=(-1)^nS\vec{z}+[(-1)^n\vec{\veps}+\vec{\eta}]\exp{i(kn-\om t)}
\eeq
with $\vec{\veps}.\vec{z}=\vec{\eta}.\vec{z}=0$. By inserting this decomposition in the equation of motion, we obtain the following dispersion relation:
\beq
\om =\pm 2JS[(\Delta-{D\over J})^2-\cos^2(ka)]^{{1\over 2}}
\eeq
In the limit $\Delta-{D\over J}\to 1^+$ we recover the same dispersion relation as  $\Delta-{D\over J}\to 1^-$ found in the planar region.

\subsection{Order by disorder along the transition lines}

In the following, we take special interest in the transition lines between these areas.
Let us first analyze the transition between the ferromagnetic and planar ground-state.
Along the line separating both regions in the phase diagram, the classical ground-state 
is {\it a priori} strongly degenerate and can be parametrized by the two spherical angles 
$(\theta,\phi)$ ($\theta$ is the angle between $\vec{S}$ and $\vec{z}$).
The Hamiltonian (\ref{hamchain}) is invariant by rotation around the $z$ axis and therefore 
the energy does not depend on $\phi$. We are here in the unusual situation where the classical
ground-state manifold is invariant under a {\em larger} group ($SO(3)$) than the Hamiltonian itself
($SO(2)$). This suggests that quantum fluctuations may reduce this larger group to the smaller one
by selecting one orbit on the classical ground-state manifold under the smaller group. In our view,
this is then a toy model for the order from disorder phenomenon.
We use the decomposition:
$$\vec{S}_n=(-1)^n\vec{S}_n^{\bot}+S_n^z \vec{z}=\vec{\s}_n^{\bot}+\vec{\s}_n^z$$
The equations of motion reduce to:
\beqa
{d \vec{\s}_n^{\bot}\over dt}&=&-J\left[(\vec{\s}_{n-1}^{\bot}+\vec{\s}_{n+1}^{\bot})\w\vec{\s}_n^z
+|\Delta|(\vec{\s}_{n-1}^{z}+\vec{\s}_{n+1}^{z})\w \vec{\s}_n^{\bot}\right]\\
 {d \vec{\s}_n^{z}\over dt}&=&-J\left[[(\vec{\s}_{n-1}^{\bot}+\vec{\s}_{n+1}^{\bot})
\w\vec{\s}_n^{\bot}\right]
\eeqa
where we have used $\Delta+{D\over J}+1=0$. We are looking for solutions around
$\vec{\s}_0=S(\cos\theta \vec{z}+\sin\theta \vec{u})$ with $\vec{u}.\vec{u}=1$ and $\vec{u}.\vec{z}=0$. 
We therefore write $\vec{\s}=\vec{\s}_0+\de \vec{\s}$ where $\de \vec{\s}$ can be parametrized by: 
$$\de \vec{\s}=[A\vec{v}+B(\cos\theta \vec{u}-\sin\theta \vec{z})]\exp{i(kn-\om t)}$$
with $\vec{v}=\vec{z}\w\vec{u}$.
After standard manipulations we find:
\beq
\label{spectre}
\om_{\theta}(k)=\pm 2JS[\cos(ka)-1][1+(|\Delta|-1)\sin^2\theta]^{{1\over 2}}
\eeq
with a dispersion in $k^2$ for the slow mode at $ka=0$.
We see that when $|\Delta|\ne 1$ there is a selection mechanism. When $|\Delta|>1$, the spin wave 
zero-point motion favours a ferromagnetic ground-state ($\theta=0$) whereas for $|\Delta|<1$ the 
planar ground-state is favoured ($\theta={\pi\over 2})$. 
 
This is the most simple example of order from disorder phenomenon in so far as the variable 
$\theta$ has no dynamics ($S^z=S\cos\theta$ is a constant of motion) and plays rather the role 
of a parameter than a true variable. If we try to classify various possible situations, we may say
here that we have a dynamical system with one conserved quantity and a 2-dimensional ground-state
manifold. Furthermore, in the present model, this ground-state manifold is generated by two
symplectic flows which do not commute (for instance the rotations along $\vec{z}$ and the rotations
along $\vec{x}$). A general model for this is given by the Hamiltonian:
\beq
H= \frac{1}{2} \sum_{i \geq 2}^{N} \sum_{j \geq 2}^{N} a_{ij}(P_{1})P_{i}P_{j}+
b_{ij}(P_{1})X_{i}X_{j}+ 2c_{ij}(P_{1})P_{i}X_{j}
\eeq
For this Hamiltonian, $P_{1}$ is conserved and therefore plays the role of
$\cos \theta$. The ground-state manifold is given by $P_{i}=X_{i}=0$ for 
$2 \leq i \leq N$. This $2D$ manifold appears as a collection of $1D$ orbits
under the symmetry generated by $P_{1}$, and characterized precisely by the
value of $P_{1}$. Note that the drift motion along these orbits arises from
the excitations of the ``fast'' variables ($2 \leq i \leq N$), as may be seen 
from the equation giving the time-derivative of $X_{1}$. Apart from this drift
motion, we can clearly separate the fast variables from the collective ones ($i=1$).

Returning to our example, it is straightforward to describe the average drift motion
$<\dot{\phi}>$ due to the quantum zero-point motion of the internal (i.e. $k \neq 0$)
modes. The idea is, as explained before, to view the zero-point motion energy of the internal
modes (which depends on $\theta$) as an effective potential for the collective variables
$\phi$ and $\theta$. Summing up the zero-point motion energy obtained from \ref{spectre}
over all $k$'s yields the effective potential:

\beq
V_{eff}(\theta)= 2 J S \sqrt{ (1+{D\over J} \sin^2 \theta)}
\eeq

This effective potential for $S^z$ leads to an average drift for the $\phi$ variable:
\beq 
\dot{\phi}={D\cos\theta\over \sqrt{1+{D\over J}\sin^2\theta}}
\eeq
Let us now consider the Ising-like case $ |\Delta| > 1$, 
for which $\theta =0$ is selected
The most interesting fact is that $\dot{\phi}$ still has a non zero limit when $\theta\to 0$. Indeed 
\beq
\label{gap}
\lim_{\theta\rightarrow 0} \dot{\phi}=D
\eeq
How shall we interprete this fact ? It simply means that the rotor has 
a residual energy even when $\theta\to 0$ which corresponds to an
easy-axis ferromagnetism. 
This is nothing  but an anisotropy gap of the system. It is easy  to check 
using Holstein-Primakoff's representation that 
the lowest excitation corresponds to a magnon with a gap $D$ for $ka=\pm \pi$. 
This simple calculation gives a classical image for a gap which has however 
a purely quantum origin, through zero-point motion of internal modes.
The results (\ref{spectre})(\ref{gap}) show that 
classical hamiltonian dynamics provides in this case a simple way to understand
the low-lying excitations of the quantum system. This analysis suggests that the critical 
theory controlling 
the behavior of the ferromagnetic chain can be described by a simple rotor 
(independently of the nature of the spin $S$). This has been analyzed in detail
and from a different point of view in \cite{rotateur}.
\vskip 0.2cm
We can perfom the same work for the transition line separating the planar 
ground-state and the AF one. The details of the algebra have been gathered 
in appendix A. The low energy spectrum is given by equation (\ref{reldis}): 

\beq
\label{reldisp}
\om _{\theta}(k)=\pm 2JS(ka)\sqrt{1+{D\over J}\sin^2\theta}
\eeq
with ${D\over J}=\D-1$. For $\D=1$, the spectrum does not depend on $\theta$. 
When $\D<1$, we find that spin waves favour the planar ground-state whereas 
for $\D>1$, the AF ground-state is selected. We have therefore the same kind of 
order from disorder phenomenon as for the Planar-Ferromagnetic transition.\\
\vskip 0.2cm
Through this example, we have analyzed a simple one dimensional example of order 
from disorder. 
Similar situations may occur in $2D$, frustrated systems like for example 
in the $J_1-J_2$ XY or Heisenberg model on a square lattice \cite{henley, simon},
where the classical degeneracy is {\em global}, and {\em partially} enforced
by a true symmetry. This list is 
obviously not exhaustive and many other  models with 
frustrating interactions experience this kind of order from disorder.
However, note that the spin chain analyzed here is special
(and actually simpler) in that the classical ground-state manifold is {\em not}
isotropic for the symplectic form which defines the classical dynamics (see section 5.1)

The Hamiltonian (\ref{hamchain}) has been studied only from the point of view of order from disorder 
using a semi-classical analysis. A full quantum analysis demands a more careful study for instance of
tunneling processes which are sensitive to the actual value of the spin (integer or half-integer). 
This is not the purpose of this article. We refer the reader to \cite{denijs} for a careful study of the phase diagram and transitions.

\section{The Husimi tree}
\setcounter{equation}{0}
\noindent
In this section, we will see another non-trivial example of the  order from disorder phenomenon, 
which we shall study in an extensive way.
We consider a pseudo-lattice called the Husimi cactus. A three-generation 
cactus has been represented on Figure 3. It consists of a succession of 
triangles connected only by their vertices. The geometrical dual of the Husimi 
cactus is a Cayley tree with constant coordination number. The interesting fact 
is that at least locally  this pseudo-lattice looks like the 
Kagom\'e lattice. However, by contrast to the Kagom\'e lattice, there is no closed 
loop of connected triangles. Each node of the cactus has one Heisenberg spin. 
We consider only nearest neighbour AF interactions, namely the Hamiltonian reads:
\beq
\label{hamcac}
H=\sum\limits_{<i,j>} \vec{S_i}\vec{S_j}
\eeq
with $\vec{S_i}^2=1$. An essential feature of the Husimi cactus is the huge degeneracy 
of its classical ground-states. This is because minimizing the 
energy on each triangular plaquette yields the constraint: 
\beq
\sum\limits_{i\in \Delta} \vec{S_i}=\vec{0}
\eeq
which does not define a unique ground-state configuration. Indeeed, 
suppose one spin on a given plaquette is fixed (say $\vec{S_1}$), then its two 
neighbours have to live 
on a half-cone centered on $-\vec{S_1}$ with an opening angle of ${\pi\over 3}$.
The ground-state manifold of a finite cactus is then a smooth manifold
(i.e. the number of zero modes is independent of the ground-state configuration).
To describe this manifold, let us assume that the top spin is fixed.
Then, we have to choose a rotation angle for the two ``offspring'' spins
around this ancestor. Because of the hierarchical structure of this lattice,
a similar freedom exists at each successive level of the hierarchy, until we reach
the bottom boundary triangles  which we shall call the leaves. Of course, we have 
a complete freedom to choose the value of the top spin, which adds two more global
zero modes. For a p-generation cactus 
with $2^p-1$ sites, the phase space manifold for the Heisenberg 
model is a product of $2^p-1$ spheres $S^{2}$, whereas the ground-state
manifold is isomorphic to the product of $S^{2}$ (choice of the top spin value)
by a torus of dimension $(2^{(p-1)}-1)$ (i.e. one angle per spin which is not on 
the bottom boundary).
Hence the ratio
\beq
{ {\rm Number~ of~ zero~ modes}\over {\rm Number~ of~ degrees~ of~ freedom}}=
{2^{(p-1)}+1\over 2(2^p-1)}\longrightarrow {1\over 4}, ~~p>>1.
\eeq

The  Husimi cactus with Heisenberg spins is interesting  so far as 
it exhibits a macroscopic number of zero modes and we expect 
spin waves to be very soft. Hence, it is not {\it a priori} obvious whether 
order from disorder can occur in such a system. In particular, are coplanar ground-states 
favoured  as in the Heisenberg Kagom\'e antiferromagnets? We emphasize however that
the degeneracy lifting mechanism is slightly different for ``real'' systems
such as the Kagom\'e or the pyrochlore lattices since for them, the most relevant
effect amounts to maximizing the number of zero modes  \cite{chalker97}. 

\subsection{Thermal fluctuations}
As we explained in the Introduction, we should distinguish the question of 
degeneracy lifting within the ground-state manifold with the more global
question of actual selection or freezing in the vicinity of a particular
ground-state. Regarding the first question, it is easy to show that classical
thermal fluctuations do not lift any degeneracy. The reason has been given in
the Introduction: a rotation of all the spins below a given site on the cactus 
around this ancestor spin preserves the phase space volume element (since the 
Jacobian matrix is then triangular and rotations preserve volumes), and also
the total energy. Therefore such transformations preserve the Gibbs measure.
Clearly any two ground-state configurations may be connected by a finite
number of such transformations. Our statement then follows easily.
Although very simple, this point has not been noticed in ref.\cite{husim}.
This paper emphasized the other aspect, namely that energy barriers
are too small to allow freezing (for instance after a quench from high 
temperatures) in any of the classical ground-state configurations.
Consequently, we do not expect an order from disorder phenomenon induced by 
thermal fluctuations in this model. This system is too soft for a selection mechanism 
to work. Other situations with extensive entropy and inequivalent ground-states, 
where there is a partial degeneracy lifting within the ground-state manifold but 
the global selection mechanism fails, have been presented in ref. \cite{husim}.
More recently, the study of the Heisenberg model on a pyrochlore lattice led 
to the same conclusions \cite{chalker97,canals}. Here, maximizing the number of zero modes
favours colinear magnetic states, but the effective barriers are too weak for the
system to remain confined in the vicinity of these states.

How similar is the quantum case?

\subsection{Quantum fluctuations}
\subsubsection{General considerations}
As in the previous sections, we wish to understand the classical spin dynamics at $T=0$ 
on the Husimi cactus. Again, the main feature of this sytem is its huge degeneracy 
and the fact that these ground-states build up a smooth manifold.
If we examine the vicinity of the ground-state manifold, we see
essentially two kinds of motion, those associated to finite frequency 
oscillations around the chosen ground-state and those associated to what 
we may call slow drift motions. Typically, this can be described by
an approximate Hamiltonian (defined around the chosen ground-state) 
which takes the general form \cite{arnold2}:   
\beq
\label{hamg}
H={1\over 2}(P_1^2+\cdots+P_r^2)+{1\over 2}\sum\limits_{j=r+s+1}^n \omega_j(P_j^2+Q_j^2)
\eeq
The $(P_i,Q_i)$, $1\le i\le r+s$ are associated to the slow
drift motions and  $(P_j,Q_j)$,\\ $r+s+1\le i\le n$  describe the 
``fast'' oscillators. Note that the $(P_i,Q_i),~r+1\le i \le r+s$ are 
associated to global symmetries. The equations of motion are:
\beqa
\label{motiong}
\dot{Q}_j=P_j&;& \dot{P}_j=0,~~1\le j\le r\nn\\
\dot{Q}_j=0&;& \dot{P}_j=0,~~r+1 \le j\le r+s\\
\dot{Q}_j=\om_j P_j&;& \dot{P}_j=-\om_jQ_j,~~r+s+1 \le j\le n\nn
\eeqa
The degenerate manifold is generated by the vectors associated to
$Q_1,\cdots,Q_{r+s};P_{r+1},\cdots,P_{r+s}$.

Our main goal is to determine if the spectrum 
depends on the ground-state we consider and more precisely if 
zero-point fluctations lift at least partially the classical degeneracy.
In order to compute the oscillator spectrum, we will proceed as follow:

(i) First we have to find in  phase space the submanifold 
associated to the drift motions. They are defined by initial 
conditions producing a speed parallel to the ground-state submanifold. 
For the general Hamiltonian (\ref{hamg}), the drift space is generated 
by the directions associated to $(P_i,Q_i)$ $1\le i\le r+s$.

(ii) Second, the oscillator submanifold is obtained by taking the symplectic 
orthogonal of the drift motions submanifold.

By symplectic form, we mean the antisymmetric bilinear form
which is defined by ${\cal G}(X,X')= \sum_{i=1}^{n}(P_{i}Q'_{i}-Q_{i}P'_{i})$,
where $X \equiv (P_{i},Q_{i})$ and  $X' \equiv (P'_{i},Q'_{i})$
are any two vectors in the $2n$-dimensional classical phase-space.
The main interest of this symplectic form is that the Poisson bracket
of two functions $f$ and $g$ defined on phase-space is given by:

$$\{f,g\}= {\cal G}(\nabla f,\nabla g).$$
This bilinear form is non-degenerate
(i.e. it is impossible to find any non-vanishing vector which is 
orthogonal to any other vector), so for any linear subspace $D$, the
set $D^{\perp}$ of all the vectors orthogonal to any vector of $D$
has a dimension $dim \: D^{\perp}= 2n-dim \: D$. We shall call
$D^{\perp}$ the symplectic orthogonal of $D$. These notions will be quite
useful for the analysis given in section 5. But it turns out that even for
basic computations, they provide substantial simplification as we shall
show shortly.

In the Husimi cactus case, phase-space is simply $(S^2)^N$, 
where $S^2$ refers to the sphere and $N$ is the total number 
of spins. If we consider only one spin $\vec{n}$ and two tangent 
vectors $d\vec{n}_1$ and $d\vec{n_2}$  to $\vec{n}$, the symplectic form ${\cal G}$ is defined by
\beq
\label{defsym}
{\cal G}(d\vec{n}_1,d\vec{n_2})=-\vec{n}.(d\vec{n}_1\w d\vec{n_2})
\eeq
It simply corresponds to the area of the parallelogram $(d\vec{n}_1,d\vec{n_2})$.
When we have more than one site, this definition extends in a straighforward way
\beq
\label{defsy}
{\cal G}({d\vec{n}},{\delta\vec{n}})=-\sum\limits_i\vec{n}_i.(d\vec{n}_i\w \de\vec{n}_i)
\eeq
where $({d\vec{n}},{\delta\vec{n}})$ are tangent vectors in phase space.
The equations of motion for this spin system are:
\beq
\label{motion}
 {d \vec{n}_{i}\over dt}= \sum_{a,i \in \D ^{a}}(\sum_{j \in \D ^{a}}\vec{n}_{j})
\w \vec{n}_{i}
\eeq
Here $\D ^{a}$ denotes any triangle on the lattice. With the present form, it
is easy to derive the linearized equations of motion in the vicinity of 
a given ground-state configuration $\vec{n}_{i}$. The spin at site $i$ is
$\vec{n}_{i}+\delta\vec{n}_{i}$, where $\vec{n}_{i}.\delta\vec{n}_{i}=0$.
For an equilibrium state, we have $\sum_{j \in \D ^{a}} \vec{n}_{j}=0$
for any triangle $\D ^{a}$. Therefore, the linearized equations read:
\beq
\label{lmotion}
{d \delta\vec{n}_{i}\over dt}= \sum_{a,i \in \D ^{a}}(\sum_{j \in \D ^{a}}\delta\vec{n}_{j})
\w \vec{n}_{i}
\eeq
Our goal is now to extract the finite frequency spectrum from these equations.
Rather than remaining abstract, let us apply this strategy to a $3$-generation cactus.

\subsubsection{Spectrum of the $3$-generation cactus}

We consider a $3$-generation cactus whose sites have been 
numbered from $1$ to $7$ (see Figure 3).  At each site, we define 
a Heisenberg spin $\vec{n_i}$. Let us derive the most general drift 
motion around an equilibrium configuration ${\vec {n_{i}}}_{1 \leq i \leq 7}$.
We have then $\vec{n_{1}}+\vec{n_{2}}+\vec{n_{3}}= 
\vec{n_{2}}+\vec{n_{4}}+\vec{n_{5}}=
\vec{n_{3}}+\vec{n_{6}}+\vec{n_{7}}=0$. This motion
is generated by a global rotation around a 
vector $\vec{r}$, a global rotation of vector $\eps \vec{n_1}$, 
a partial rotation of $\vec{n_4}$ and $\vec{n_5}$ around $\vec{n_2}$ 
(of angle $\eta$), and finally a partial rotation of $\vec{n_6}$ 
and $\vec{n_7}$ around $\vec{n_3}$ (of angle $\zeta$).
We are looking for tangent vectors ${\vec{\de n_i}}$ ($\vec{n_i}.\de \vec{ n_i}=0$) 
producing a motion parallel to the ground-state manifold. Therefore, 
we have to solve the system (the arrows on vectors will 
be omitted in order to lighten expressions): 
\beq
\label{sw1}
\left\{ \begin{array}{l}
\dot{n}_1=r\w n_1=(\de n_1+\de n_2+ \de n_3)\w n_1\\ 
\dot{n}_2=(r+\eps n_1)\w n_2=(\de n_1+\de n_2+ \de n_3+\de n_2 +\de n_4  +\de n_5)\w n_2\\ 
\dot{n}_3=(r+\eps n_1)\w n_3=(\de n_1+\de n_2+ \de n_3+\de n_3 +\de n_6  +\de n_7)\w n_3 \\
\dot{n}_4=(r+\eps n_1+\eta n_2)\w n_4=(\de n_2+ \de n_4+\de n_5 )\w n_4 \\  
\dot{n}_5=(r+\eps n_1+\eta n_2)\w n_5=(\de n_2+ \de n_4+\de n_5 )\w n_5 \\  
\dot{n}_6=(r+\eps n_1+\zeta n_3)\w n_6=(\de n_3+ \de n_6+\de n_7 )\w n_6 \\  
\dot{n}_7=(r+\eps n_1+\zeta n_3)\w n_7=(\de n_3+ \de n_6+\de n_7 )\w n_7 
\end{array}\right.
\eeq
The fact that we are in the vicinity of an equilibrium state 
imposes that $n_5$ and $n_6$ are linearly independent 
as well as $n_6$ and $n_7$. This fact is used in the resolution of the 
system (\ref{sw1}) and we thus find that the submanifold 
associated to the drift motion is locally defined by:
\beq
\label{espace1}
\left\{ \begin{array}{l}
\de n_1+\de n_2+ \de n_3=0\\ 
\de n_2+ \de n_4+\de n_5=\eps n_1+\eta n_2\\  
\de n_3+ \de n_6+\de n_7=\eps n_1+\zeta n_3
\end{array}\right.
\eeq
We call $D_1$ this subspace. The case $\eps=\eta=\zeta=0$ 
corresponds to initial conditions with a vanishing speed, 
namely to the tangent space to the degenerate submanifold,
which we shall denote by $D_{2}$.
We have now to take the symplectic orthogonal of this submanifold $D_1$. 
More accurately, we are looking for vectors $u_i\in (D_1)^{\bot}$ such that
\beqa
\label{cond}
i)~~&&\forall i ,~u_i.n_i=0 \nn\\
ii)~&& \forall~ {\de n_i} \in D_1,~ \sum_i n_i.(\de n_i\w u_i)=0
\eeqa
The procedure to construct explicitly this submanifold is 
not straighforward and is explained in Appendix B. Nevertheless, 
in order to determine the spectrum, we may work with the larger
submanifold $(D_2)^{\bot}$ which has been defined in 
appendix B. To give some illustration of what these subspaces mean,
let us return to the ``flat'' canonical phase-space used in the context
of the general Hamiltonian \ref{hamg}. The drift space $D_{1}$ is as
we have said generated by the vectors associated to $(P_{i},Q_{i})$ for
$1 \leq i \leq r+s$, and $D_{1}^{\bot}$ corresponds to the finite
frequency modes $(P_{i},Q_{i})$, $r+s+1 \leq i \leq n$. The space
$D_{2}$ is spanned by the vectors associated to $Q_{i}$, $1 \leq i \leq r+s$
and $P_{j}$, $r+1 \leq j \leq r+s$. Its orthogonal $D_{2}^{\bot}$ contains
the direct sum of $D_{1}^{\bot}$ and the subspace of $D_{2}$ generated by
the vectors associated to $Q_{i}$, $1 \leq i \leq r$. As the equations of motion
\ref{motiong} show, $D_{2}^{\bot}$ is stable for the Hamiltonian flow.
The velocity vanishes everywhere on $D_{2}^{\bot} \bigcap D_{2}$, and the image
of  $D_{2}^{\bot}$ for the linear mapping $(P_{i},Q_{i}) \rightarrow 
(\dot{P}_{i},\dot{Q}_{i})$ is precisely the subspace associated to the finite
frequency modes, namely $D_{1}^{\bot}$.

Consider now a vector $u\in D_2^{\bot}$ which can be defined by:
\beq
\label{compuspace}
\left\{\begin{array}{c}
u_1=u_1^a;~u_2=u_2^a+u_2^b;~ u_3=u_3^a+u_3^c\\ 
u_4=u_4^b;~u_5=u_5^b;~u_6=u_6^c;~u_7=u_7^c\end{array}\right.
\eeq
with $u_1^a+u_2^a+u_3^a=u_2^b+u_4^b+u_5^b=u_3^c+u_6^c+u_7^c=0$.

We now compute the flow associated to $u$ which is naturally defined as
\beq
\label{flot}
\left\{\begin{array}{l}
v_1=(u_1+u_2+u_3)\w n_1=(u_2^b+u_3^c)\w n_1=v_1^a\\  
v_2=(u_1+u_2+u_3+u_2+u_4+u_5)\w n_2=(u_2^a+u_2^b+u_3^c)\w n_2=v_2^a+v_2^b\\  
v_3=(u_1+u_2+u_3+u_3+u_6+u_7)\w n_3=(u_3^a+u_3^c+u_2^b)\w n_3=v_3^a+v_3^c\\  
v_4=(u_4+u_5+u_2)\w n_4=u_2^a\w n_4= v_4^b\\  
v_5=(u_4+u_5+u_2)\w n_5=u_2^a\w n_5= v_5^b\\  
v_6=(u_3+u_6+u_7)\w n_6=u_3^c\w n_6= v_6^c\\  
v_7=(u_3+u_6+u_7)\w n_7=u_3^c\w n_7= v_7^c
\end{array}\right.
\eeq
where the velocities $(v_{i}=\dot{n}_{i})$ are given by the linearized
equations of motion \ref{lmotion}.
By defining
\beq
\left\{\begin{array}{l}
v_1^a=(u_2^b+u_3^c)\w n_1;~v_2^a=(u_2^b+u_3^c)\w n_2;~v_3^a=(u_3^c+u_2^b)\w n_3\\  
v_2^b=u_2^a\w n_2;~ v_4^b=u_2^a\w n_4;~v_5^b=u_2^a\w n_5\\  
v_3^c=u_3^a\w n_3;~ v_6^c=u_3^a\w n_6;~v_7^c=u_3^a\w n_7\end{array}\right.\eeq
we check that $(D_2)^{\bot}$ is stable under the Hamiltonian flow. Indeed, we clearly
have:
$$v^{a}_{1}+v^{a}_{2}+v^{a}_{3}=
v^{b}_{2}+v^{b}_{4}+v^{b}_{5}=v^{c}_{3}+v^{c}_{6}+v^{c}_{7}=0.$$
Moreover, we can prove that $v\in (D_1)^{\bot}$, as we discussed using the
Hamiltonian (\ref{hamg}). 

To show explicitly this 
interesting property here, we have just to show that $v$ is orthogonal to 
any tangent vectors $\de n$ satisfying (\ref{espace1}). 
The demonstration proceeds in a pedestrian way. The main ingredient involved is that
${\cal G}(v,\de n)$ involves only the sums of the $n$ field on the various
triangles, which are precisely given by the right-hand side of equations
\ref{espace1}. A similar reasoning is detailed at the end of Appendix B.
 
We now compute the frequencies of the oscillators. We therefore consider the square of the application giving the Hamiltonian flow:
\beq
\label{square}
u\stackrel{\rm flow}{\longrightarrow} v\stackrel{\rm flow}{\longrightarrow} -\om^2 u \eeq
The application closes because of the stability of $D_2^{\bot}$ under the linearized flow.
We obtain:
\beqa
&&\left\{\begin{array}{c}
-\om ^2 u_1^a=(v_2^b+v_3^c)\w n_1=(u_2^a\w n_2+u_3^a\w n_3)\w n_1\\
-\om ^2 u_2^a=(v_2^b+v_3^c)\w n_2=(u_2^a\w n_2+u_3^a\w n_3)\w n_2\\
-\om ^2 u_3^a=(v_2^b+v_3^c)\w n_3=(u_2^a\w n_2+u_3^a\w n_3)\w n_3\end{array}\right.\\
&&\left\{\begin{array}{c}
-\om ^2 u_2^b=v_2^a\w n_2=((u_2^b+u_3^c)\w n_2)\w n_2\\
-\om ^2 u_4^b=v_2^a\w n_4=((u_2^b+u_3^c)\w n_2)\w n_4\\
-\om ^2 u_5^b=v_2^a\w n_5=((u_2^b+u_3^c)\w n_2)\w n_5
\end{array}\right.\\
&&\left\{\begin{array}{c}
-\om ^2 u_3^c=v_3^a\w n_3=((u_2^b+u_3^c)\w n_3)\w n_3\\
-\om ^2 u_6^c=v_3^a\w n_6=((u_2^b+u_3^c)\w n_3)\w n_6\\
-\om ^2 u_7^c=v_3^a\w n_7=((u_2^b+u_3^c)\w n_3)\w n_7\end{array}\right.
\eeqa
We notice that the equations associated to $(u_2^a,u_3^a)$  and to $(u_2^b,u_3^c)$ are closed. 
We have therefore to solve two sytems
\beqa
&&(A)\left\{\begin{array}{c}
-\om ^2 u_2^a=-u_2^a+(u_3^a.n_2)n_3+{1\over 2} u_3^a\\
-\om ^2 u_3^a=-u_3^a+(u_2^a.n_3)n_2+{1\over 2} u_2^a\end{array}\right.\\ &&\nn\\
&&(B)\left\{\begin{array}{c}
-\om ^2 u_2^b=-(u_2^b+u_3^c)+((u_2^b+u_3^c).n_2).n_2\\
-\om ^2 u_3^c=-(u_2^b+u_3^c)+((u_2^b+u_3^c).n_3).n_3\end{array}\right.
\eeqa
In order to solve the system $(A)$, we introduce  the vector $z={n_1\w n_2\over ||n_1\w n_2||}$ and
write $$u_i^a=\rho_i z+\s z\w n_i,~~1\le i\le 3,$$ and $$\rho_{1}+\rho_{2}+\rho_{3}=0$$.
We find for system $(A)$ that the possible values for $\om ^2$ are 
${1\over 2},~{3\over 2},~2$. We can solve the system $(B)$ in the same 
coordinate system and we find the same eigenvalues.

We can thus conclude that the eigenvalues of the massive modes for the $3$-generation 
cactus are $\om ^2={1\over 2},~\om ^2={3\over 2},~\om ^2=2$ and are doubly degenerate.

Moreover, the spectrum {\bf does not} depend on the classical ground-state and the 
order from disorder phenomenon does not occur for this system.

\subsubsection{Order by disorder in  the $p$-generation cactus}

We would like to see now if these results also apply to any finite cactus with
$p$ generations. The general structure of the calculations is as
outlined in the previous paragraph. We also work within the linear space
$D_{2}^{\bot}$ whose definition is the natural generalisation of equations \ref{compuspace}.
Since $D_{2}^{\bot}$ is stable under the linearized flow, we define a linear
mapping of $D_{2}^{\bot}$ which sends any tangent vector ${u_{i}}$ of 
$D_{2}^{\bot}$ into the corresponding velocity ${v_{i}}$. An explicit calculation
of the secular determinant for the $p=4$ cactus is presented in Appendix C.
The main conclusion is that for $p \geq 4$, the zero-point energy does depend 
on the classical ground-state configuration. An illustration of this is
given on Figure 5 where the zero-point energy is plotted along a special curve
on the classical ground-state manifold. This curve corresponds to the ground-states
which originate from a coplanar configuration and are then obtained by rotating all
the spins below site 2 (i.e sites 2,4,5,8,9,10,11) by an angle $\theta\in [0,\pi]$ around
$\vec{n}_{2}$. The zero-point energy is minimal for $\theta = \pi k$, ($k$ integer) and maximal
for $\theta = (k+\dmi)\pi$. Therefore, zero-point fluctuations favour coplanar 
states, and induce effective barriers between such states. We remark that all the
planar ground-states are equivalent even after introducing zero-point fluctuations.
This is because the Husimi cactus has a huge space-group which preserves both the
Hamiltonian and the symplectic structure. For instance, consider
one triangle  $\al,\beta,\ga$ in the bulk of the cactus 
($\al$ being the top site of the triangle). If we exchange the 
sites $\be$ and $\ga$ and also all their descendants, it is clear the lattice is
preserved. It is also clear that any coplanar ground-state may be mapped into any
other by a finite sequence of such transformations. Now, Figure 5 also
suggests that these degenerate coplanar states are connected by rather low effective
energy barriers. Indeed, the height of the barrier is about $1\%$ of the low energy excitation which is very low. We would have thought intuitively to a much higher barrier
since we rotate all spin below sites $2$. It indicates that the effective low energy landscape is rather flat and that tunelling processes will play a very important role.
On intuitive grounds, we expect these effective barriers to remain {\it finite} as the system size goes to infinity, since such a rotation around a given ancestor spin induces only {\it local} corrections to the small oscillation matrix (see appendix E3).


Another feature of the cactus is that this degeneracy lifting mechanism occurs 
{\em only} in the bulk of the cactus but not on the leaves.
The $3$-generation cactus appears in fact as a particular case because it 
is built  only of leaves (the leaves are defined as the bottom boundary triangles). 
We have demonstrated these results for the $4$-generation tree, but 
the conclusion still holds for a $p$-generation cactus ($p>4$) 
by a straightforward induction proof. The special case of leaves is addressed in
section 5.3. 

This set of results suggests that physical properties of this system will be
dominated at $T=0$ by dramatic tunneling processes within the discrete set
of coplanar classical ground-states. This will be the subject of further work \cite{husim2}.

\vskip 0.2cm
 A simpler example in the same class 
is what is referred to in the litterature as the Delta chain. It consists
of a chain of triangles linked by their vertices as shown in Figure 6. This system has been much studied and the specific heat was shown to have a double peak structure as in the Kagom\'e antiferromagnet \cite{kubo93}. Therefore, it was believed that  an understanding of its properties can shed light on the Kagom\'e antiferromagnet. Theoretical interests on this chain have also been enhanced  by its experimental realization \cite{walstedt}. Many studies have been devoted to 
 its extreme quantum version
namely for $S=1/2$ spins \cite{kubo93,kubo96,nakamura96}. The ground-state is exactly known to be a dimer state \cite{suto}. The low lying excitations were found to be kink and antikink-type domain walls in the dimer singlet ground-state \cite{kubo96}.

Our semi-classical analyses applies directly to an {\em open} Delta chain, and we find that coplanar ground-states minimize the spin-wave zero-point energy here as well.

\section{A geometrical analysis of the spin dynamics on the Husimi cactus}
\setcounter{equation}{0}
\subsection{Properties of the flow associated to rotations around a given spin}

Let us consider one spin at a site labelled  $i$. The infinitesimal rotation of angle $\eps$ around this spin defines a vector field on phase-space. More accurately, if the phase space is $(S^2)^N$ (with $N$ the number of spins), a tangent vector  $\vec{m}_i$ to the configuration $\vec{n}_i$ verifies $\vec{m}_i.\vec{n}_i=0$   ($i\in [1,\cdots,N]$).
If $j$ is a descendant of $i$ then $\vec{m}_j=\eps\vec{n_i}\w\vec{n}_j$ else $\vec{m}_j=0$.
We will characterize and analyze in the following the properties of these vector fields.

\begin{itemize}
\item The first and main property of these vector fields is that the energy is conserved along their flow in  phase-space.

\item Secondly, we may wonder if these vector fields can be associated to the symplectic flow of a fonction defined in phase-space (that would therefore commute with the Hamiltonian!).
According to the previous section, we know that the answer to this issue is negative because a particular order is selected (indeed, if the hypothesis were true, it would imply that actual symmetry protects these internal  rotations and therefore prevents any selection mechanism from acting). Nevertheless, we now prove it in a direct way. In a general manner, suppose we have a vector field $U^i$ on a manifold and $g_{ij}$ denotes the symplectic form in any coordinate system. We assume that:
\beq \exists F ~{\rm such~ that}~U^i=g^{ij}\p_jF
\eeq
It implies that $\forall i,~\al_i=g_{ij}U^j=\p_iF$, where $\al_i$ is a one-form.
We consider the simple case of two spins $\vec{n}_1$
and $\vec{n}_2$ and take the rotation of  $\vec{n}_2$ around $\vec{n}_1$ of angle $\eps$.
This defines a vector field $\vec{m}_1=0;~\vec{m}_2=\eps\vec{n_1}\w\vec{n}_2$.
If $d\vec{n_1}$ and $d\vec{n}_2$ are respectively tangent vectors to $\vec{n_1}$ and $\vec{n}_2$ then: 
\beqa
\label{exform}
\al_1.d\vec{n}_1+\al_2.d\vec{n}_2&=&-\vec{n}_1.(d\vec{n}_1\w\vec{m}_1)-\vec{n}_2.(d\vec{n}_2\w\vec{m}_2)\nn\\&=&-\eps\vec{n}_2.(d\vec{n}_2\w (\vec{n}_1\w\vec{n}_2))=\vec{n}_1.d\vec{n}_2
\eeqa
This property comes from $$\al_i.d\vec{n}^i=d\vec{n}^ig_{ij}U^j={\cal G}(d\vec{n}_1,d\vec{n}_2)$$ where ${\cal G}(.,.)$ has been defined in (\ref{defsym}).
From (\ref{exform}), it appears that $\al_id\vec{n}^i$ is not  an exact differential because it is not closed. Indeed, the equivalent of the rotational is defined by
\beq\label{defb}\p_i\al_j-\p_j\al_i=b_{ij},\eeq
If $d\vec{n}_i$ and $\de\vec{n}_i$ ($i\in[1,2])$ are two tangent vectors to the point  $\vec{n}_i$, then 
\beq
\label{defba}b_{ij}dn^i\de n^j=d\vec{n}_1\de\vec{n}_2-\de\vec{n}_1d\vec{n}_2\ne 0\eeq
We can therefore conclude that those flows associated to internal rotations around a given spin are not symplectic flows of  fonctions defined on the phase-space despite they do not change the energy.

\item Another interesting property that may be checked concerns the commutativity of the flows associated to two different rotations. Suppose, we first perform a rotation around a spin $i$ and then a rotation around a spin $j$ which is a descendant of spin $i$. This is not {\it a priori} obvious that the two operations commute. However, we have shown  in the appendix D that this is the case.

\item Finally, we want to mention another property of  vector fields. Let us consider two internal rotations and their two associated vector fields say $U$ and $V$. Suppose $U$ corresponds to a rotation of angle $\eps$ around one spin $\vec{n}_i$ and $V$ to a rotation of angle $\eta$ around one spin $\vec{n}_j$. If $i$ and $j$ are not comparable in the lattice hierarchy then ${\cal G}(U,V)=0$. But if $j$ is a descendant of $i$, then ${\cal G}(U,V)$ has no reason to be equal to $0$.

Nevertheless, on the ground-state manifold, it is easy to verify that
\beq \forall U,\forall V,~~{\cal G}_{\big| ~{\rm Ground-State}}(U,V)=0\eeq
This follows directly from the fact the ground-state property $\sum\limits_i \vec{S}_i^{(a)}$, $i$ indicating the three sites of a given triangle $a$.
In more technical words, the submanifold of the ground-state manifold generated  by these internal rotations is isotropic for the symplectic form.

\end{itemize}

\subsection{Vector flow associated to internal rotations and Hamiltonian Flow}
In this section, we want to clarify why the flow associated with these internal rotations does not in general commute with the Hamiltonian flow (except for the rotations of the leaves of the Husimi tree). As we shall see, this is closely related to the non-vanishing of the $2-$form $b_{ij}$ (\ref{defba}).

\subsubsection{General considerations}

In this subsection, we derive some general considerations which will be useful for our problem. We consider a phase-space with a symplectic closed form ${\cal G}$ (of components $g_{ij}$), a Hamiltonian H which is invariant under the flow associated to a vector field $X$. It reads as:
\beq 
\label{rela1}{\cal L}_{X}H=X^i\p_i H=0,\eeq with ${\cal L}_{X}$ the Lie derivative in the $X$ direction. We also define $Y$ to be  the Hamiltonian vector field which therefore satisfies 
\beq 
\label{rela2}Y^i=g^{ij}\p_j H\eeq
The quantity we are interested in is the Lie bracket:
\beq
\label{rela2b}
[X,Y]^i=X^j\p_j Y^i-Y^j\p_j X^i
\eeq

\no By using the definitions (\ref{rela2},\ref{rela2b}) and the invariance of $H$ under $X$ (\ref{rela1}), we can easily prove that:
\beqa
\label{rela3}
[X,Y]^i&=&-\left\{-(\p_j g^{ik})X^j+g^{ij}(\p_jX^k)+g^{jk}(\p_jX^i)\right\}(\p_k H)\nn\\
&=& -a^{ik}\p_k H
\eeqa
where we have defined $a^{ik}=-(\p_j g^{ik})X^j+g^{ij}(\p_jX^k)+g^{jk}(\p_jX^i)$.

\no The relation (\ref{rela3}) can also be formulated as:
\beq
[X,Y]^i=-g^{ij}a_{jk}Y^k
\eeq
with:
\beq
\label{defa}
a_{ij}=g_{ik}a^{kl}g_{lj}=-a_{ji}=(\p_k g_{ij})X^k+g_{ik}(\p_j X^k)+g_{kj}(\p_iX^k)
\eeq
We can therefore conclude 
that: \beq
\label{equiv}
[X,Y]=0 \equi a_{ij}Y^j=0
\eeq
The $2-$form $a_{ij}$ plays a crucial role.
We first restrict to the simplest hypothesis $\forall i,j~a_{ij}=0$. 
This condition has a simple interpretation in differential geometry and means that $X$ is a Killing vector field \cite{geom}. It means that the symplectic form $g_{ij}$ is invariant under the flow associated to $X$, {\it i.e.}

$${\cal L}_{X}{\cal G}=0$$
or in components:
\beq
(\p_k g_{ij})X^k+g_{ik}(\p_j X^k)+g_{kj}(\p_iX^k)=0
\eeq

\no Therefore 
$a_{ij}=0\equi {\cal L}_{X}{\cal G}=0$. Moreover, if $X$ is a Killing vector field, it is 
at least locally the flow associated to a conserved quantity $F$ \cite{geom} (a conserved quantity because the flow of $X$ leaves $H$ invariant \footnote{
To show that, we suppose $X^i=g^{ij}\p_jF\equi \p_iF=g_{ij}X^j$. By using the Schwartz equality $\p_i\p_j F=\p_j\p_iF$ and the Leibnitz rule, we prove the equivalence. Note that this also  requires that $g_{ij}$ is closed {\it i. e.} $\p_ig_{jk}+\p_jg_{ki}+\p_kg_{ij}=0$. }.  

From these generalities, we can conclude that a {\em sufficient} condition for the flow $U$ associated to internal rotations to commute with the Hamiltonian flow is that $U$ should be the flow of a conserved quantity. We have proved in section 5.1, that it cannot be true for our symplectic form.

Nevertheless, we must not forget that the condition (\ref{equiv}) is weaker. In the Husimi cactus, we have $a_{ij}\ne 0$ as already suggested by eq. (\ref{defba}) and  as will be seen further. A natural issue occurs, namely can we find a vector field $X$ leaving the Hamiltonian invariant such that  $a_{ij}\ne 0$ but $a_{ij}Y^j=0$ ?
Before analysing this question, let us make a pause to summarize the important consequences of the equivalences derived above. 

The relation (\ref{equiv}) implies the following statement:

 {\it In a system with a huge continuous classical degeneracy, the order from disorder phenomenon (induced by quantum fluctuations) will in almost  cases occur unless these degeneracies are protected by true symmetries of the Hamiltonian}.
Note that here, order from disorder refers to degeneracy lifting within the ground-state manifold and not to the {\it global} behaviour of the system.

\subsubsection{Application to the Husimi cactus}

We now turn back to the question where $a_{ij}\ne 0$. Can we find in the Husimi cactus vector fields such that $a_{ij}Y^j=0$ ?
Suppose we consider the vector field $X$ associated to the rotation of angle $\eps$ around $\vec{n}_i$ of the vectors $\vec{n}_j$ ($\vec{n}_j$ are descendant spins of $\vec{n}_i$).
The $2-$form   $a_{ij}$ is also given by $\p_i(g_{jk}X^k)-\p_j(g_{ik}X^k)$ which  corresponds to the $2-$form $b_{ij}$ defined in (\ref{defb}). Therefore, if $\vec{u}_j$ and $\vec{v}_j$ refer to two tangent vectors to the point $\vec{n_j}$, then
according to relation (\ref{defba}),
\beq
\label{defauv}
a(u,v)=\sum\limits_{j\ge i} \vec{u}_i.\vec{v}_j-\vec{u}_j.\vec{v}_i
\eeq
where the notation $j\ge i$ means $j$ descendant of $i$.
We now consider the $1$-form $\be_i=a_{ij}V^j$, where $V^j$ is the Hamiltonian flow at the point $\vec{n}_j$ in phase-space. For the Husimi tree, it is defined by: 
\beq
\vec{v}_j=\sum\limits_{k\in N_j}\vec{n}_k\w\vec{n}_j
\eeq
where $N_j$ means the nearest neighbours sites of $j$.
Consequently,
\beq
\be(u)\equiv a(u,v)=\sum\limits_{j\ge i}\left\{\sum\limits_{k\in N_j}(\vec{n}_k\w\vec{n}_j)\right\}\vec{u}_i- \sum\limits_{k\in N_i}(\vec{n}_k\w\vec{n}_i).
\sum\limits_{j\ge i}\vec{u}_j
\eeq
This summation has absolutely no reason to vanish. We can thus conclude according to (\ref{equiv}) that the Hamiltonian flow  does not commute with the flow associated to internal rotations (except the case of the leaves of the Husimi tree as we will see in the next subsection). To analyze this phenomenon more precisely, we shall also consider the vicinity of an equilibrium state, and study the evolution of the small oscillation matrix  under the action of the flow $X$ associated to internal rotations. This has been developed extensively in appendix E. We only summarize the main results.
First, we have recalled in appendix E1  how the variation of the oscillation matrix $A$ under the flow $X$ is connected to the $2-$form $a_{ij}$:

\beqa
&&({\cal L}_X A)_{ij}=0\equi-a_{ik}A^k_{~j}=0\\
&&\equi \forall ~u,v~~a(u,Av)=0
\eeqa
It is another way of showing that the low-energy spectrum is invariant with the flow of $X$ when $X$ is a Killing vector field.

 We have then applied this general formulation 
to the Husimi cactus in appendix E2  and proved explicitly that the oscillation matrix is not invariant under the flow associated to internal rotations (see eq:(\ref{aw})). 

And finally, in appendix E3, we have compared the oscillation matrices for two configurations connected by an internal rotation and shed light on the general dependence of the low energy spectrum with the angle of the rotations.

\subsection{Case of the leaves of the Husimi tree}

We now treat the problem of the leaves of the tree. We consider one leaf $(\vec{n}_0,\vec{n}_1,\vec{n}_2)$ with $(\vec{n}_1,\vec{n}_2)$ the two edge spins. We have seen in appendix B that the degeneracy still holds. According to our geometrical discussion in subsection 5.1, this degeneracy must therefore be associated to a conserved quantity.

For one triangle, the Hamiltonian simply reduces to 
\beq
H=(\vec{n}_0+\vec{n}_1+\vec{n}_2)^2\eeq
This degeneracy still occurs because the quantity $(\vec{n}_1+\vec{n}_2)^2$ commutes with the Hamiltonian (under Poisson brackets). Indeed, the symplectic flow associated to 
$(\vec{n}_1+\vec{n}_2)^2$  is given by
\beq
\left\{ \begin{array}{l}
{d\vec{n}_1\over d\la}=\vec{n}_2\w\vec{n}_1\\  \\
{d\vec{n}_2\over d\la}=\vec{n}_1\w\vec{n}_2 \\  \\
{d\vec{n}_i\over d\la}=0 ~~{\rm if}~i\not \in\{1,2\}  \end{array}\right.
\eeq
where $\la$ is the parameter associated to the transormation.
Therefore,
\beq
{d(\vec{n}_1+\vec{n}_2)\over d\la}=0~~{\rm and}~~{d H\over d\la}=0
\eeq
It is worth noticing that such a transformation cannot be in general identified with all the rotations we may imagine, namely:

\no
-a rotation around $\vec{n}_0$\\
-a rotation around $\vec{n}_0^0$ excluding  $\vec{n}_0$\\
-a rotation around $\vec{n}_0^0$ including  $\vec{n}_0$\\

\no Here, $\vec{n}_i^0$ denotes an equilibrium value, and we assume $\vec{n}_i$  is close to $\vec{n}_i^0$.
The only exception corresponds to $(\vec{n}_1+\vec{n}_2)$ colinear to  $\vec{n}_0$ for the first case and to  $\vec{n}_0^0$ for the last two cases.

Furthermore, we can prove explicitly that the order from disorder phenomenon does not apply for the leaves of the cactus. We will simply check in the following that the flow associated to the function $f=(\vec{n}_1+\vec{n}_2)^2$ commutes with the Hamiltonian flow.
If we replace in Figure 7, the application $R$ by $f$, we have therefore to show (using the same notations) that $\xi=\tilde{\xi}$.

We consider an infinitesimal transformation associated to the fonction $f=(\vec{n}_1+\vec{n}_2)^2$.
Therefore,
$$(\vec{n}_1;\vec{n}_2)\lra(\vec{n}_1+\eps \vec{n}_2\w \vec{n}_1;\vec{n}_2+\eps \vec{n}_1\w \vec{n}_2)$$
Let us consider a tangent vector $(\vec{\eta}_1,\vec{\eta}_2)$ to $(\vec{n}_1;\vec{n}_2)$ in phase-space. The symplectic transformation $f$ implies that:
\beq
(\vec{n}_1+\vec{\eta}_1;\vec{n}_2+\vec{\eta}_2)\lra(\vec{n}_1+\vec{\eta}_1+\eps\vec{n}_2\w \vec{n}_1+\eps \vec{n}_2\w \vec{\eta}_1+\eps\vec{\eta}_2\w \vec{n}_1;\vec{n}_2+\vec{\eta}_2+\eps \vec{n}_1\w \vec{n}_2+\eps \vec{n}_1\w \vec{\eta}_2+\eps\vec{\eta}_1\w \vec{n}_2)
\eeq
Therefore
\beq
(\vec{\eta}_1;\vec{\eta}_2)\lra (\vec{\psi}_1;\vec{\psi}_2)=(\vec{\eta}_1+\eps \vec{n}_2\w \vec{\eta}_1+\eps\vec{\eta}_2\w \vec{n}_1;\vec{n}_2+\eps \vec{n}_1\w \vec{\eta}_2+\eps\vec{\eta}_1\w \vec{n}_2)
\eeq
The Hamiltonian flow in the initial configuration defines the $\vec{\xi}_i$ $i\in\{0,1,2\}$
as follows
\beq
\left\{\begin{array}{l}
\vec{\xi}_0=-\vec{n}_0^0\w[\vec{\eta}_0+\vec{\eta}_1+\vec{\eta}_2+\vec{\eta}_0+\vec{\eta}_{-1}+\vec{\eta}_{-2}]\\
\vec{\xi}_1=-\vec{n}_1^0\w[\vec{\eta}_0+\vec{\eta}_1+\vec{\eta}_2]\\
\vec{\xi}_2=-\vec{n}_2^0\w[\vec{\eta}_0+\vec{\eta}_1+\vec{\eta}_2]\end{array}\right.
\eeq
where the sites $(-1)$ and $(-2)$ are the NN of site $(0)$, in the ancestor triangle.

Because of $\vec{\psi}_0=\vec{\eta}_0$ and $\vec{\psi}_1+\vec{\psi}_2=\vec{\eta}_1+\vec{\eta}_2$, we have
$$\vec{\eta}_T=[\vec{\eta}_0+\vec{\eta}_1+\vec{\eta}_2]=\vec{\psi}_T=[\vec{\psi}_0+\vec{\psi}_1+\vec{\psi}_2]$$.

Therefore, following the notations of Figure 7,
\beq
\left\{\begin{array}{l}
\vec{\chi}_0=-\vec{n}_0^0\w[\vec{\eta}_T+\vec{\eta}_0+\vec{\eta}_{-1}+\vec{\eta}_{-2}]\\
\vec{\chi}_1=-[\vec{n}_1^0+\eps\vec{n}_2^0\w \vec{n}_1^0]\w\vec{\eta}_T\\
\vec{\chi}_2=-[\vec{n}_2^0+\eps\vec{n}_1^0\w \vec{n}_2^0]\w\vec{\eta}_T
\end{array}\right.
\eeq
By applying the inverse transformation of $f$ to the vectors $\chi_i$, we can easily prove that
\beq
\forall j~~\tilde{\xi}_j=f^{-1}(\vec{\chi}_j)=\xi_j
\eeq
 Therefore, the existence of a true symetry (associated to $(\vec{n}_1+\vec{n}_2)^2$) leads to the absence of order from disorder for all the spins belonging to the edge of the Husimi cactus. Note that it is in complete agreement with the geometric vision of order from disorder developed along this paper.

\section{Summary,   Conclusions and perspectives}
\setcounter{equation}{0}

Our main conclusions have been in fact already presented in the Introduction.
The main result of this work is to show that continuous degeneracies which are
not enforced by continuous symmetries of the Hamiltonian are more fragile
quantum-mechanically than under the action of classical thermal fluctuations. This
is because classical thermal fluctuations are not sensitive to the underlying
Poisson bracket structure which is of course crucial for quantization. To put
this work in perspestive, we give in a Table \ref{tab} a summary of what we believe is the
present status of three closely related systems: the Kagom\'e, the pyrochlore 
lattices and the Husimi cactus. The first line of the Table \ref{tab} refers to the local
analysis of degeneracy lifting. We have included both thermal and quantum fluctuations.
Note that the degeneracy lifting mechanism is quite different between the ``real'' lattices
and the Husimi cactus. For the former, optimal states are the ones which maximize the number 
of zero modes. This generates a discrete subset of classical ground-states (namely the 
coplanar states for the Kagom\'e lattice and the colinear states for the pyrochlore lattice),
within which a finer selection may occur (for instance favouring an ordered state on the
Kagom\'e lattice). Note that the zero mode counting mechanism does not introduce much difference between classical thermal fluctuations and quantum zero-point fluctuations.
However, the residual finer selection is sensitive to whether the system is classical or quantum mechanical. This mechanism is of course not available on the Husimi cactus, for
which the number of zero modes is independent of the ground-state configuration. The second line of the table summarizes the global selection effects namely whether the system remains or not  close to any of the favoured ground-states.
 
Two extensions of the present work may be considered. The first one would be to
set-up a similar semi-classical analysis for systems in which the number of zero
modes actually depends on the ground-state configuration. As we have just mentioned, 
this would be more relevant to ``real'' systems such as the Kagom\'e or the pyrochlore
lattices. The other direction would be to take advantage of the simpler
phase-space structure on the cactus to analyze tunneling processes within the
discrete set of coplanar ground-states. As usual, these effects are sensitive for 
instance to the value of the quantum spin $S$ \cite{henley92}, and a rich pattern
of possible quantum phases as $S$ varies may occur in such a system. Numerical
diagonalizations would therefore be very instructive, specially if performed for
several values of the spin $S$.

\vspace*{2cm}
\noindent
{\bf Acknowledgements:} We would like to thank P. Azaria and F. Mila for
several interesting discussions concerning this work.
\vspace*{2cm}

\newpage

\begin{tabular}{||l|l|l||}
\hline\hline
 &Classical Husimi cactus& Quantum Husimi cactus \\
\hline
Local degeneracy &No&Yes (coplanar) \\
lifting& {\it this work} & {\it this work}\\
\hline
Global & No &No \\ 
selection&\cite{husim} &\cite{husim} \\ \hline\hline
\end{tabular}
\vskip 2cm

\begin{tabular}{||l|l|l||}
\hline\hline
& Classical Kagom\'e &Quantum kagom\'e \\
\hline
Local degeneracy&Coplanar & Ordered  \\
lifting&
or ordered \cite{chalker92} & \cite{sachdev}-\cite{chandra93} \\ \hline
Global & Coplanar states  & No (Spin Liquid)\\ 
selection&\cite{chalker92} & \cite{lechem} \\ \hline\hline
\end{tabular}

\vskip 2cm
\begin{table}[h]
\begin{tabular}{||l|l|l||}
\hline\hline
&Classical Pyrochlore&Quantum Pyrochlore \\
\hline
Local degeneracy&Colinear state & Colinear state  \\
lifting&
 \cite{reimers} &  \\ \hline
Global &No   & No (Spin Liquid)\\ 
selection&\cite{chalker97} & \cite{chalker97,canals} \\ \hline\hline
\end{tabular}

\vspace*{1cm}
\caption{Status of the ground-states of the classical and quantum Heisenberg model on the  Husimi, Kagom\'e and pyrochlore lattices with respect to local or global selection phenomena}
\label{tab}
\end{table}

\newpage 
\appendix
\section{Low energy spectrum along the Planar-AF transition  }
\setcounter{equation}{0}
We suppose ${D\over J}-\Delta+1=0$ with $\D>0$. Because of the AF behaviour, we use the decomposition $\vec{S}_n=(-1)^n\vec{\s}_n+a\vec{l}_n$ with $\vec{\s}_n.\vec{l}_n=0$. $a$ represents the lattice constant. Because of the anisotropy in the $z$ direction, we are forced to separate $\vec{\s}_n$ and $\vec{l}_n$ in two parts:
$$\vec{\s}_n=\vec{\s}_n^{\bot}+\vec{\s}_n^{z}~~;~~\vec{l}_n=\vec{l}_n^{\bot}+\vec{l}_n^z$$
We can write the equations satisfied by these four vectors (we  directly take the  continuum limit):
\beqa
{d\vec{\s}^{\bot}\over dt}&=&4aJ(\vec{l}^{\bot}\w\vec{\s}^{z})+Da^3(\vec{l}^{\bot}\w {\p^2
\vec{\s}^{z}\over \p x^2})+ 4a(D+J) (\vec{l}^{z}\w\vec{\s}^{\bot})\\
{d\vec{\s}^{z}\over dt}&=&4aJ (\vec{l}^{\bot}\w\vec{\s}^{\bot})\\
{d\vec{l}^{\bot}\over dt}&=&-(D+J)a^2({\p^2\vec{\s}^{z}\over \p x^2}\w\vec{\s}^{\bot})
-Ja^2({\p^2\vec{\s}^{\bot}\over \p x^2}\w\vec{\s}^{z})\\
{d\vec{l}^{z}\over dt}&=&-Ja^2 ({\p^2\vec{\s}^{\bot}\over \p x^2}\w\vec{\s}^{\bot})
\eeqa
As in section 2, we are looking for solutions around $\vec{S}^0=S(\cos\theta \vec{z}+\sin\theta \vec{u})$ with $\vec{u}.\vec{u}=1$ and $\vec{u}.\vec{z}=0$.
We write $\vec{S}=\vec{S}^0+\de \vec{S}$ where $\de \vec{S}=(-1)^n\de\vec{\s}+a\vec{l}$.
Using the above equations we obtain the following system for $\de \vec{\s}^z,\de\vec{\s}^{\bot},\vec{l}^{z},\vec{l}^{\bot}$:

\beqa
\label{sy1}
{d\de\vec{\s}^{\bot}\over dt}&=&4aJS(\vec{l}^{\bot}\w\cos\theta\vec{z})+ 4a(D+J)S (\vec{l}^{z}\w\vec{u}\sin\theta)\\
{d\de\vec{\s}^{z}\over dt}&=&4aJ S(\vec{l}^{\bot}\w\cos\theta\vec{z})\\
{d\vec{l}^{\bot}\over dt}&=&-(D+J)a^2S \sin\theta({\p^2\de\vec{\s}^{z}\over \p x^2}\w\vec{u})
-Ja^2S\cos\theta({\p^2\de\vec{\s}^{\bot}\over \p x^2}\w\vec{z})\\
\label{sy2}
{d\vec{l}^{z}\over dt}&=&-Ja^2 S  \sin\theta({\p^2\de\vec{\s}^{\bot}\over \p x^2}\w\vec{u})
\eeqa
We use the following form for $\de\vec{\s}$ and $\vec{l}$:
\beqa
\de\vec{\s}&=&\left[A \vec{z}\w\vec{u}+B\vec{w}\right]\exp{i(kx-\om t)} \nn\\
\vec{l}&=&\left[C \vec{z}\w\vec{u}+E\vec{w}\right]\exp{i(kx-\om t)}
\eeqa
where $\vec{w}=-\sin\theta \vec{z}+\cos\theta\vec{u}$. Using the basis $(\vec{z},\vec{u},\vec{v}=\vec{z}\w\vec{u})$, and the system (\ref{sy1}-\ref{sy2}), we obtain four equations: \eject
\beqa
i\om B&=&-4aJC\\
i\om A&=&4E[J+D\sin^2\theta]\\
i\om E&=&-J(ka)^2A\\
i\om C&=&(J+D\sin^2\theta)(ka)^2B
\eeqa
The resolution of this simple system gives the dispersion relation
\beq
\label{reldis}
\om =\pm 2JS(ka)\sqrt{1+{D\over J}\sin^2\theta}
\eeq
with ${D\over J}=\D-1$.

\section{Vectors associated to the spectrum for the $3$-generation cactus}
\setcounter{equation}{0}
In this appendix, we are looking for vectors  $u_i\in (D_1)^{\bot}$ such that:
\beqa
\label{defu1}
i)~~&&\forall i ,~u_i.n_i=0 \nn\\
ii)~&& \forall~ {\de n_i} \in D_1,~ \sum_i n_i.(\de n_i\w u_i)=0
\eeqa
where $D_1$ is defined by (\ref{espace1}). 
We first determine the  symplectic orthogonal of the subspace 
$D_2$ defined by the solutions of the homogeneous equations:
\beqa
\label{espace2}
\de n_1+\de n_2+ \de n_3&=&0\nn\\
\de n_2+ \de n_4+\de n_5&=&0\\
\de n_3+ \de n_6+\de n_7&=&0\nn
\eeqa
We have obviously $D_2 \subset D_1$. Therefore the symplectic 
orthogonal of $D_1$ is included in the one of $D_2$. It is easier 
to work with the submanifold $D_2$ since $D_2$ is defined by an 
intersection of subspaces defined by only one condition: 
$$D_2^a=\{{\de n_{i}}  / \sum\limits_{i\in \Delta^a} \de n_i^a=0\}$$
where $a$ indicates the $a^{th}$ triangle and $i\in \{1,2,3\}$. 
Hence $D_2=\bigcap\limits_{a} D_2^a$. Therefore the symplectic orthogonal 
of $D_2$ is given by:
$$D_2^{\bot}=\bigoplus\limits_a (D_2^a)^{\bot}$$ since the 
symplectic form is non-degenerate. This elementary algebra 
has interesting consequences because we have only to work on 
one triangle. 
Indeed, the first thing to notice is that a vector ${u_{i}}$ tangent  to an
equilibrium configuration ${n_{i}}$ and belonging to $(D_2^a)^{\bot}$
vanishes on all the sites which do not belong to the triangle labelled by $a$.
To show this, we remark that any tangent vector ${\de n_{i}}$ for which
$\de n^{a}_{1}= \de n^{a}_{2} = \de n^{a}_{3} =0$ belongs to $D^{a}_{2}$.
So ${u_{i}}$ has to be orthogonal to all such ${\de n_{i}}$'s.
More precisely, this implies $\sum_{i \not\in \D ^{a}} n_i.(\de n_i\w u_i)=0$
for any set of ${\de n_{i}}$ defined for $i \not\in \D ^{a}$ (and of
course $\de n_{i}.n_{i}=0$). Since $u_{i}.n_{i}=0$, this implies that
$u_{i}=0$ for any $i \not\in \D ^{a}$. Therefore, we have only to determine
$u_{i}$ for $i \in \D ^{a}$. For this, it is convenient to consider a 
unit vector $z^{a}$ normal to the plane containing $(n^{a}_{1},n^{a}_{2},n^{a}_{3})$
(note that $\sum_{i=1}^{3} n^{a}_{i}=0$). The tangent vectors $\de n_{i}$
and $u_{i}$ to $n_{i}$ restricted to the plaquette $\D ^{a}$ are then conveniently
expressed as:
\begin{eqnarray}
u_{i} &=& \rho _{i} z^{a} + \sigma _{i} (z^{a} \w n^{a}_{i}) \\
\de n^{a}_{i} &=& \lambda _{i} z^{a} + \mu _{i} (z^{a} \w n^{a}_{i})
\end{eqnarray}
After an elementary algebra (using $n^{a}_{i}.n^{a}_{j}=-1/2$ for $i \neq j$),
we find that $D^{a}_{2}$ restricted to $\D ^{a}$ is the set of tangent vectors
$\de n^{a}_{i}$ such that $\lambda_{1}+\lambda_{2}+\lambda_{3}=0$ and
$\mu_{1}=\mu_{2}=\mu_{3}$. This can be shown to imply that
$\rho_{1}+\rho_{2}+\rho_{3}=0$ and $\sigma_{1}=\sigma_{2}=\sigma_{3}$. 
So $(D^{a}_{2})^{\bot}$ is the subspace of $D^{a}_{2}$ of tangent vectors
which vanish everywhere but on the triangle $\D ^{a}$. To summarize, and
changing notation slightly, for each triangle $\D ^{a}$ containing the sites
$\al(a),\be(a),\ga(a)$, the tangent vectors
${u_i^a}$ in $(D^{a}_{2})^{\bot}$
are defined by:
\beqa
\label{defu}
&&i)~~ \forall i,~~u_i^a.n_i=0\nn\\
&&ii)~~u_i^a=0~~\forall i \ne \al(a),\be(a),\ga(a)\\
&&iii)~~ u_{\al(a)}^a+u_{\be(a)}^a+u_{\ga(a)}^a=0\nn
\eeqa

One vector of $D_2^{\bot}$ is written naturally as $u_i=\sum\limits_a u_i^a$.

In order to form $D_1^{\bot}$, we have just to impose the orthogonality to the particular solutions of the inhomogeneous equations (\ref{espace1}). 
The symplectic form of a vector $u$ in $D_2^{\bot}$ with a tangent vector $\{\de n\}$ equals:
\beq
{\cal G}[u,\de n]=-\sum\limits_{a} \sum\limits_i n_i.(u_i^a \w \de n_i)
\eeq
According to (\ref{defu}), $u_i^a=0$ if $i$ does not belong to the triangle $a$, we can therefore focus on one particular triangle $a$. With the same notations as before, the contribution of the triangle $a$ to the symplectic form reads as:
\beq
{\cal G}[u^a,\de n]=-\s(\la_1+\la_2+\la_3)+\rho_1\mu_1+\rho_2\mu_2+\rho_3\mu_3
\eeq
It is then easy to show that the symplectic form ${\cal S}[u^a,\de n]$ depends {\it only} on the sum $\de n_1+\de n_2+\de n_3$
\beq
{\cal G}[u^a,\de n]=(\de n_1^a+\de n_2^a+\de n_3^a).\left[ {2\over 3}\sum\limits_i u_i^a \w n_i^a+ \s z^a\right]
\eeq
The main point is that the sum  $\de n_1+\de n_2+\de n_3$ 
 is just the quantity involved in the equations (\ref{espace1}).
The dimension of $(D_2)^{\bot}$ is $9$. For $(D_1)^{\bot}$, we have three more conditions. Its dimension is therefore $6$ and we have consequently only three independent massive modes.
More explicitly, for the $3$-generation cactus (see Figure 3 for notations),  $(D_1)^{\bot}$ is parametrized by:
\beqa
\s^a,~\s^b,~\s^c&&\nn\\
\rho_1^a+\rho_2^a+\rho_3^a&=&0\nn\\
\rho_2^b+\rho_4^b+\rho_5^b&=&0\nn\\
\rho_3^c+\rho_6^c+\rho_7^c&=&0
\eeqa
and they are constrained by: 
\beqa
&&n_1.\left[{2\over 3}z^b \w (\rho_2^b n_2+\rho_4^b n_4+\rho_5^b n_5)-\s^b z^b\right]+\nn\\
&&~~~~~~~~~~~~~~~~~~~~~n_1.\left[
{2\over 3}z^c \w (\rho_3^c n_3+\rho_6^c n_6+\rho_7^c n_7)-\s^c z^c\right]=0\nn\\
&&~~~~~~~~~~~~~~~~~~~~~n_2.\left[{2\over 3}z^b \w (\rho_2^b n_2+\rho_4^b n_4+\rho_5^b n_5)-\s^b z^b\right] =0\\
&&~~~~~~~~~~~~~~~~~~~~~n_3.\left[
{2\over 3}z^c \w (\rho_3^c n_3+\rho_6^c n_6+\rho_7^c n_7)-\s^c z^c\right]=0\nn
\eeqa
 
\section{Vectors associated to the spectrum for the $4$-generation cactus}
\setcounter{equation}{0}

We consider the $4$-generation cactus whose sites and triangles have been labelled following Figure $4$. We follow exactly the same construction as for the $3$-generation cactus.
We thus define vectors $u_i^{\al}$ $i\in [1,..,15];~\al \in \{a,..,g\}$ such that 
\beqa
&&u_i^{\al}.n_i=0\nn\\
&&u_i^{\al}=0~~{\rm if~ i~ is~ not~ a~ site~ belonging~ to~ the~ \al^{th}~ triangle}\\
&&\sum\limits_i~ u_i^{\al}=0, ~~i {\rm ~is~ a~ site~ of~ the ~ \al^{th}~ thiangle}\nn
\eeqa
The vectors $u$ define the symplectic orthogonal $D_2^{\bot}$ of the degenerate submanifold $D_2$. The subspace $D_2^{\bot}$ is stable under the linearized hamiltonian flow. This defines vectors $v_i^{\al}$ as: 
\beqa
v_j^a&=&(u_2^b+u_3^c)\w n_j~,~j\in\{1,2,3\}\nn\\
v_j^b&=&(u_2^a+u_4^d+u_5^e)\w n_j~,~j\in\{2,4,5\}\nn\\
v_j^c&=&(u_3^a+u_6^f+u_7^g)\w n_j~,~j\in\{3,6,7\}\nn\\
v_j^d&=&(u_4^b)\w n_j~,~j\in\{4,7,9\}\\
v_j^e&=&(u_5^b)\w n_j~,~j\in\{5,10,11\}\nn\\
v_j^f&=&(u_6^c)\w n_j~,~j\in\{6,12,13\}\nn\\
v_j^g&=&(u_7^c)\w n_j~,~j\in\{7,14,15\}\nn
\eeqa
As in the previous case, we can show that the vectors $v$ are in fact in the submanifold $D_1^{\bot}$. In order to find the eigenvalues of the flow associated to the vectors $u$, we consider the square of the linearized hamiltonian flow application. We have therefore the following system to solve:
\beqa
\label{sys1}
-\om^2 u_j^a&=&\left[(u_2^a+u_4^d+u_5^e)\w n_2+(u_3^a+u_6^f+u_7^g)\w n_3\right]\w n_j,~j\in \{1,2,3\}\\
\label{sys2}
-\om^2 u_j^b&=&\left[(u_2^b+u_3^c)\w n_2+u_4^b\w n_4+ u_5^b\w n_5\right]\w n_j,~j\in \{2,4,5\}\\
\label{sys3}
-\om^2 u_j^c&=&\left[(u_2^b+u_3^c)\w n_3+u_6^c\w n_6+ u_7^c\w n_7\right]\w n_j,~j\in \{3,6,7\}\\
\label{sys4}
-\om^2 u_j^d&=&\left[(u_2^a+u_4^d+u_5^e)\w n_4\right]\w n_j,~j\in \{4,8,9\}\\
\label{sys5}
-\om^2 u_j^e&=&\left[(u_2^a+u_4^d+u_5^e)\w n_5\right]\w n_j,~j\in \{5,10,11\}\\
\label{sys6}
-\om^2 u_j^f&=&\left[(u_3^a+u_6^f+u_7^g)\w n_6\right]\w n_j,~j\in \{6,12,13\}\\
\label{sys7}
-\om^2 u_j^g&=&\left[(u_3^a+u_6^f+u_7^g)\w n_7\right]\w n_j,~j\in \{7,14,15\}
\eeqa
Let us notice a rather surprising fact: this sytem divides into two subsystems of independent variables. The first subset contains equations (\ref{sys2},\ref{sys3}) associated  to triangles $(b),~(c)$, the second subsystem contains equations (\ref{sys1}),(\ref{sys3}-\ref{sys7}). Therefore, these equations connect next nearest neighbour triangles to each other. This property is clearly general and does not depend upon the number of generations of the Husimi cactus. This is associated to the fact that we consider the square of the Hamiltonian flow application. 

We first consider the systems of equations (\ref{sys2},\ref{sys3}) related to triangle $(b)$ and $(c)$. We define a system of coordinates where:
$$n_1=(1,0,0)~;~n_2=(-{1\over 2},{\sqrt{3}\over 2},0)~;~n_3=(-{\sqrt{3}\over 2},-{1\over 2},0),~$$
where $z^a=(0,0,1)$ is a vector orthogonal to the plane containing $n_1,n_2,n_3$ and is linked to triangle $(a)$. In a similar manner, we can define vectors $z^b,~z^c$ associated to triangles $(b)$ and $(c)$. The vectors $u$ decompose on this basis as follows:
\beqa
u_i^b&=&\rho_i^bz^b+\s^b (z^b\w n_i^b)~,~ i\in\{2,4,5\}~~{\rm with}~~ \sum\limits_i\rho_i^b=0\nn\\
u_j^c&=&\rho_j^c z^c+\s^c (z^c\w n_j^b)~,~j \in\{3,6,7\} ~~{\rm with}~~ \sum\limits_j\rho_j^c=0
\eeqa
We first consider the equation (\ref{sys2}) with $j=2$:
\beq
\label{u2}
\om^2 u_2^b-(u_2^b+u_3^c)+n_2.(u_2^b.n_2+u_3^c.n_2)+{1\over 2}(u_4^b+u_5^b)+n_4.(u_4^b.n_2)+n_5.(u_5^b.n_2)=0
\eeq
We then use the properties of the vectors $u_i$, namely $u_2^b+u_4^b+u_5^b=0$ and $u_2^b.n_2=0$. We project this equation on the $z\w n_2$ axis and use the relations:
\beqa
&&u_4^b.n_2=-{\sqrt{3}\over 2}\s^b\nn\\
&&u_5^b.n_2= {\sqrt{3}\over 2}\s^b\nn
\eeqa
The vectors $z^b$ and $z^c$ can be parametrised respectively with the angles $\theta_b$ and $\theta_c$ which represent the angles made with $z_a$. Therefore, we have
$$z^b=\left(\eps^b{\sqrt{3}\over 2}\sin(\theta_b),\eps^b{1\over 2}\sin(\theta_b),\cos(\theta_b)\right)$$
$$z^c=\left(\eps^c{\sqrt{3}\over 2}\sin(\theta_c),-\eps^c{1\over 2}\sin(\theta_c),\cos(\theta_c)\right)$$
with $\eps^b,\eps^c=\pm 1$.
After projection on the $z\w n_2$ axis, the equation (\ref{u2}) becomes:
\beqa
\label{w1}
(\om^2-3)\s^b&+&\s^c\left[{1\over 2}\cos\theta_b\cos\theta_c+\eps^b\eps^c\sin\theta_b\sin\theta_c\right]\nn\\&+&\rho_3^c\left[{1\over 2}\eps^c\cos\theta_b\sin\theta_c-\eps^b\sin\theta_b\cos\theta_c\right]=0
\eeqa
Using similar algebra, we project (\ref{u2}) on the $z^b$  axis and obtain
\beqa
\label{w2}
(\om^2-{3\over 2})\rho_2^b&-&\rho_3^c\left[\cos\theta_b\cos\theta_c+\eps^b\eps^c\sin\theta_b\sin\theta_c\right]\nn\\
&+&\s^c\left[\eps^c\cos\theta_b\sin\theta_c-{\eps^b\over 2}\sin\theta_b\cos\theta_c\right]=0
\eeqa
If we  consider  equation (\ref{sys2}) with $j=4,5$, we obtain equations compatible with
equations (\ref{w1},\ref{w2}).
We can treat the system (\ref{sys3}) in a similar way and we obtain in this case:
\beqa
\label{w3}
(\om^2-3)\s^c&+&\s^b\left[{1\over 2}\cos\theta_b\cos\theta_c+\eps^b\eps^c\sin\theta_b\sin\theta_c\right]\nn\\&+&\rho_2^b\left[{1\over 2}\eps^b\cos\theta_c\sin\theta_b-\eps^c\sin\theta_c\cos\theta_b\right]=0
\eeqa
\beqa
\label{w4}
(\om^2-{3\over 2})\rho_3^c&-&\rho_2^b\left[\cos\theta_b\cos\theta_c+\eps^b\eps^c\sin\theta_b\sin\theta_c\right]\nn\\
&+&\s^b\left[\eps^b\cos\theta_c\sin\theta_b-{\eps^c\over    2}\sin\theta_c \cos\theta_b\right]=0
\eeqa
In the coplanar case ($\theta_b=\theta_c=0$), we obtain the following eigenvalues for $\om^2$: ${1\over 2},~{5\over 2},~{7\over 2}$ with double degeneracy. When ($\theta_b,\theta_c\ne 0$), we have to compute the eigenvalues of a $4\times 4$ matrix.
The characteristic polynomial reads: 
\beqa
\label{det}
&&{\cal P}(w^2=X,\theta_1,\theta_2)=X^4-9X^3+28X^2-\left(36+{9\over 8}(\cos^2\theta_b+\cos^2\theta_c)\right)X\\
&&+{175\over 16}-{45\over 16}\sin^2\theta_b\sin^2\theta_c+{63\over 16}(\sin^2\theta_b+\sin^2\theta_c)-{9\over 4}\cos\theta_b\cos\theta_c\sin\theta_b\sin\theta_c\nn
\eeqa
and therefore depends on the ground-state !

Notice that this result is in apparent contradiction with the case of the $3$-generation cactus where the spectrum does not depend on the ground-state. In the $4$-generation cactus, a selection mechanism holds. It is easy from (\ref{det}) to see that the configuration minimizing the energy is the planar one ($\theta_b=\theta_c=0$). Under these conditions we find four eigenvalues $w^2={1\over 2}^{(1)},~{5\over 2}^{(2)},~{7\over 2}^{(1)}$. 

Therefore, the spins belonging to the triangles $(a),~(b),~(c)$ are coplanar. Rather than solving the heavy system of equations (\ref{sys1}),(\ref{sys4}-\ref{sys7}), we can use an extension of the result proved in the previous appendix. Indeed, if we consider one centered triangle $b$ connected to three free triangles by its vertices. The results of appendix B extend in a straighforward way, and no selection mechanism is found in this system with $9$ spins. Therefore, the three  pairs of spins can rotate independently. 
If we apply this result to triangles $(b),(a),(d),(e)$ ($(b)$ being the centered triangle), we find that the spins $(8,9)$ and ($10,11)$ are free to rotate though the spins of triangle $(a)$ are fixed.
 Consequently, to summarize, there is no selection mechanism in the leaves of the Husimi tree!

\section{Lie brakets between vectors fields associated to internal rotations}
\setcounter{equation}{0}
In this appendix, we check in a pedestrian way that the flow associated to two different internal rotations commute. In the Husimi tree, we consider $R_i(\theta)$ and $R_j(\eta)$,
with $R_i(\theta)$ the rotation of angle $\theta$ around $\vec{n}_i$. Three cases have to be looked upon. We have therefore to compare the result of the application $R_i(\theta)R_j(\eta)$ with $R_j(\eta)R_i(\theta)$ on a vector $\vec{n}_k$ ($k\ne i,j$). 

\begin{itemize}

\item If $i$ and $j$ are not connected by a descendence relation, the result is trivial.

\vskip 0.2cm
\item If $k$ is between $i$ and $j$, then
\beqa
R_i(\theta)R_j(\eta)(\vec{n}_k)&=&R_i(\theta)(\vec{n}_k)\nn\\
R_j(\eta)R_i(\theta)(\vec{n}_k)&=&R_i(\theta)(\vec{n}_k),
\eeqa
which ensures the result.

\vskip 0.2cm
\item If $k$ is below both $i$ and $j$, the result is {\it a priori} less trivial. 
 Let us assume that $i$ is an ancestor of $j$. We have to compare:

\beq \label{rotation}
R[\vec{n}_i,\theta ]R[\vec{n}_j,\eta)](\vec{n}_k)~~{\rm and}~~
R[R(\vec{n}_i,\theta)(\vec{n}_j);\eta]R[\vec{n}_i,\theta](\vec{n}_k)\eeq
where $R[\vec{n},\al]$ denotes the rotation of angle $\al$ around $\vec{n}$.

A little algebra shows that this property indeed holds and that the two quantities defined in (\ref{rotation}) are equal.
\end{itemize}

\section{Vicinity of the equilibrium manifold}
\setcounter{equation}{0}
\subsection{General point of view}
We first analyze the vicinity of  a sub-manifold ${\cal E}$ of equilibrium state of an Hamiltonian $H$. Therefore, ${\cal E}$ is defined by:
\beq
x\in {\cal E} \equi \left\{\begin{array}{c} H(x)=0\\
\p_iH(x)=0\end{array}\right.\eeq

Let $x_0\in {\cal E}$. We consider a small deviation around $x_0$ and thus write
$x^i=x_0^i+y^i$ with $||y||<<||x_0||$. The equation of motion for $y$ is:
\beq
{dy^i\over dt}=g^{ij}(x_0+y)\p_jH(x_0+y)
\eeq
Using $||y||<<||x_0||$, this equation becomes:
\beq
\label{mvt}
{dy^i\over dt}=g^{ij}(x_0)\p_j\p_kH(x_0)y^k+O(y^2)
\eeq
Defining $A^i_{~j}(x_0)=g^{ij}(x_0)\p_j\p_kH(x_0)$, the linearized equation of motion  around $x^0$ simply reads:
\beq
{dy^i\over dt}=A^i_{~j}(x_0)y^j
\eeq
It is worth noticing that $A^i_{~j}$ is {\it not} in general a tensor. It can be checked easily that $ \p_j\p_kH$ does not transform in a covariant way under coordinate transformations. However, if we restrict ourselves to the submanifold ${\cal E}$, the matrix 
$A^i_{~j}(x_0)$ does define a tensor.
Let us now express the invariance of $A$ under the flow associated to $X$ on the submanifold ${\cal E}$:
\beq
{\cal L}_X A=0\equi (\p_kA^i_{~j})X^k+A^i_{~k}\p_jX^k-A^k_{~j}\p_kX^i=0
\eeq
with ${\cal L}$ the Lie derivative.\\
Let us now extend this definition in the vicinity of  ${\cal E}$, though $A$ is no longer a tensor. We therefore define $M^i_{~j}= (\p_kA^i_{~j})X^k+A^i_{~k}\p_jX^k-A^k_{~j}\p_kX^i$. \\Using  $A^i_{~k}=g^{ij}\p_j\p_kH$, 
we obtain
\beq
\label{defm}
M^i_{~j}=(\p_kg^{il})X^k(\p_l\p_jH)+g^{il}X^k(\p_k\p_l\p_jH)+g^{il}(\p_jX^k)(\p_l\p_kH)-
g^{kl}(\p_kX^i)(\p_l\p_jH)
\eeq
The invariance of $H$ under the flow of $X$  is expressed by ${\cal L}_XH=0$ or in coordinates
\beqa
\label{floh}
&&X^k\p_kH=0\Lra(\p_jX^k)(\p_kH)+X^k(\p_j\p_kH)=0\\
\label{floh1}
&\Lra&(\p_j\p_lX^k)(\p_kH)+(\p_jX^k)(\p_l\p_kH)+(\p_lX^k)(\p_j\p_kH)+X^k(\p_j\p_k\p_l)H=0
\eeqa 
The term $(\p_j\p_lX^k)(\p_kH)$ equals zero on ${\cal E}$.
By injecting (\ref{floh1}) in (\ref{defm}), we find that:
\beqa
\label{relma}
M^i_{~j}&=&\left\{(\p_kg^{il})X^k-g^{ik}(\p_kX^l)-g^{kl}(\p_kX^i)\right\}(\p_l\p_jH)\nn\\
&=&-a^{il}\p_l\p_kH,
\eeqa
where $a^{ij}$ has been introduced in (\ref{rela3}). The relation (\ref{relma}) can be still written as
\beq
\label{ma}
M_{ij}=-g_{im}a^{mn}g_{nk}g^{kl}(\p_l\p_jH)=-a_{ik}A^k_{~j}
\eeq 
Not surprisingly, we recover the $2-$form $a_{ij}$ corresponding to ${\cal L}_Xg$. We can therefore conclude that if $X$ is a Killing vector field, then the matrix $A$ of small oscillations around an equilibrium configuration is invariant under the flow associated to $X$.
\subsection{Application to the Husimi cactus}

Let $\vec{u}$ and $\vec{v}$ two tangent vectors to an equilibrium configuration $\{\vec{n}^0\}$ of  phase-space.
According to (\ref{ma}),
\beq
M(\vec{u},\vec{v})=-u^ia_{ik}A^k_{~j}v^j=-a(\vec{u},A\vec{v})
\eeq
We recall how to build the matrix $A$ for the Husimi tree. We consider $\vec{n}_i=\vec{n}_i^0+\vec{v}_i$ with $\vec{n}_i.\vec{v}_i=0$.
Therefore,
\beqa
(A\vec{v})_i&=&{d\vec{n}_i\over dt}=\sum\limits_{j\in N_i}\vec{n}_j\w\vec{n}_i\nn\\
&=&\sum\limits_{j\in N_i}\vec{n}_j^0\w\vec{v}_i+\sum\limits_{j\in N_i}\vec{v}_j\w\vec{n}_i^0+O(v^2)
\eeqa
where $N_i$ indicates all the nearest neighbours of $i$. At lowest order,
\beq
\label{defz}
(A\vec{v})_i=\vec{w}_i=-\vec{n}_i^0\w\left(\sum\limits_{k\in N_i}\vec{v}_k+{1\over 2}z_i\vec{v}_i\right)\eeq
where $z_i$ is the coordination number of site $i$. Using the expression (\ref{defauv}) of $a(u,w)$, we find
\beq
\label{av}
a(\vec{u},A\vec{v})=\sum\limits_{j>i}(\vec{u}_i\vec{w}_j-\vec{u}_j\vec{w}_i)
\eeq
where $\vec{w}$ has been defined in the previous relation.
The expression $$\sum\limits_{j>i}\vec{w}_j=-\sum\limits_{j>i} \vec{n}_j^0\w\left(\sum\limits_{k\in N_j}\vec{v}_k+{1\over 2}z_j\vec{v}_j\right)$$  can be simplified by noticing that:

\begin{itemize}

\item For $k>i$ and $k\not \in N_i$, the term containing $\vec{v}_k$ vanishes because
$(\sum\limits_{j\in N_k}\vec{n}_j^0+{1\over 2}z_k\vec{n}_k^0=0$) for an equilibrium state.

\item For $k>i$ and $k\in N_i$, the term containing $\vec{v}_k$ is: 
$$-\left(\sum\limits_{j\in N_k}\vec{n}_j^0+{1\over 2}z_k\vec{n}_k^0\right)\w\vec{v}_k=
\vec{n}_i^0\w\vec{v}_k$$

\item For $k=i$, the term containing  $\vec{v}_i$ is $\vec{n}_i^0\w\vec{v}_i$
\vskip 0.2cm

\end{itemize}

\no From these three types of contributions, we infer:
\beq
\label{aw}
a(\vec{u},A\vec{v})=\vec{n}_i^0.\left\{\left(\sum\limits_{k>i,k\in N_i}\vec{v}_k+\vec{v}_i\right)\w\vec{u}_i\right\}+ \vec{n}_i^0.\left\{\left(\sum\limits_{k\in N_i} \vec{v}_k+2\vec{v}_i\right)\w
\sum\limits_{j>i}\vec{u}_j\right\}
\eeq
Consequently, the $2-$form is clearly different from zero. This result proves that rotations of spins $\vec{n}_j$ around $\vec{n}_i$ (with $j$ descendant of $i$) do not preserve the form of the oscillation matrix around equilibrium. Notice also that it explains why a direct calculation of the low-energy spectrum with standard methods is difficult to implement. However, this result does not enable us to show explicitly that the matrices $A^i_{~j}$ of two equilibrium configurations are  not equivalent. In order to prove it, we have to compute the characteristic polynomial and show it depends on  the chosen ground-state as we checked explicitly in appendix C for the $4$-generation cactus.
It seems  difficult to get general analytical expressions  as in appendix C for a $p$-generation cactus. Another  less ambitious solution consists in analyzing  how the oscillations matrices of two different equilibrium configurations are connected. This is the subject of the next subsection of this appendix.

\subsection{Links between two different classical configurations}
Consider one vector $\eta$ in phase-space.
We want to compare the Hamiltonian flow of $\eta$  with another one linked to $\eta$ by a
rotation around $\vec{n}_i^0$. Furthermore, it provides us with a way of connecting the matrices of small oscillations around two different classical ground-states. This is schematized in Figure 7. We are to compare $\xi$ and 
$\tilde{\xi}$.
In this sheme, we first begin with a state $\eta$.
The state $\psi$ is deduced by a rotation around $\vec{n}_i^0$ of angle $\theta$, therefore,
\beq
\left\{\begin{array}{c}
\psi_j=R(\vec{n}_i^0,\theta)\eta_j ~~{\rm if}~j>i\\
\psi_j=\eta_j~~{\rm elsewhere}
\end{array}\right.\eeq
We have to distinguish the cases $j>i$ and $j\ge i$. 
We first examine the Hamiltonian flow around this new equilibrium configuration. It defines a state $\chi$ as follow
\beqa
\chi_j&=&-R(\vec{n}_i^0,\theta)[\vec{n}_j^0]\w\left(\sum\limits_{k\in N_j}\psi_k+{1\over 2}z_j\psi_j\right)~~{\rm if} ~j>i\\
\chi_j&=&-\vec{n}_j^0\w\left(\sum\limits_{k\in N_j}\psi_k+{1\over 2}z_j\psi_j\right)~~
{\rm elsewhere}
\eeqa
We have used (\ref{defz}) in order to establish these equations.
We now apply the inverse rotation $R(\vec{n}_i^0,-\theta)$ (see Figure 7). We also use the property discussed in appendix D:
\beq
R(\vec{n}_i^0;-\theta)R(R(\vec{n}_i^0,\theta)[\vec{n}_j^0];\eta)R(\vec{n}_i^0;\theta)=R(\vec{n}_j^0;\eta)
\eeq
In the limit $\eta\to 0$, we deduce:
\beq
\label{prop1}
R(\vec{n}_i^0;-\theta)\left\{R(\vec{n}_i^0,\theta)[\vec{n}_j^0]\w R(\vec{n}_i^0,\theta)[\vec{u}]\right\}=\vec{n}_j^0\w\vec{u}
\eeq

 We have to distinguish four different cases:
\begin{itemize}

\item $j>i$ and $j$ has not $i$ as a nearest neighbour  (NN). In this case
$$\forall k\in N_j,~\psi_k=R(\vec{n}_i^0,\theta)\eta_k~.$$

Therefore,
\beq
\tilde{\xi}_j=-\vec{n}_j^0\w\left(\sum\limits_{k\in N_j}\eta_k+{1\over 2}z_j\eta_j\right)
=\xi_j\eeq

\item $j>i$ and $j$ NN of $i$. In this case $\psi_i=\eta_i$. Using (\ref{prop1}), we show
\beq
\tilde{\xi}_j=-\vec{n}_j^0\w\left(\sum\limits_{k\in N_j,k\ne i}\eta_k+R(\vec{n}_i^0;-\theta)\eta_i+{1\over 2}z_j\eta_j\right)
\eeq

\item $j=i$
\beq
\tilde{\xi}_j=-\vec{n}_i^0\w\left(\sum\limits_{k\in N_{j=i},k< i}\eta_k+  
 \sum\limits_{k\in N_{j=i},k> i}  R(\vec{n}_i^0;\theta)\eta_k+ {1\over 2}z_i\eta_i\right)
\eeq

Notice that the notation $k<i$ means that $k$ is not a descendant of $i$.

\item $j\ne i$ and $j<i$. In this case, we trivially show that $\tilde{\xi}_j=\xi_j$.
\end{itemize}

\vskip 0.2cm
\no We may also consider the case of the rotation of the $\vec{n}_j$'s around $\vec{n}_i^0$ for $j\ge i$, and similar conclusions can be drawn.

Consequently, we can infer from this analysis that the problem of  small oscillations around a new configuration (deduced by an internal rotation around spin $\vec{n}_i^0$) reduces to a problem of small
oscillation in the initial configuration but with a modification of the bounds between $i$
and its two immediate descendants (the new ``interaction'') depends of $\theta$. This modification seems to be the origin of the dependence of the low energy spectrum with the angle $\theta$. Note that after this change of basis, the modification of the linearized equation of motion due to a change in $\theta$ is only local. This is why we do not expect large effective energy barrier between different coplanar ground-states.

\eject

\newpage
\centerline{\bf FIGURES CAPTIONS}
\vspace*{2cm}

\no FIG.1a: Poincar\'e sections of the slow ($1$) and the fast ($2$) motions for $\eps=0.1$ with rescaling of the variables $(X_2,P_2)$.
The adiabatic approximation seems good even for rather high values of $\eps$.

\vspace*{0.5cm}
\no FIG.1b: Poincar\'e sections for the slow and fast motions for $\eps=0.25$. We begin to see the deviations from the circular orbits signaling violation of the adiabatic approximation.

\vspace*{0.5cm}
\no FIG.1c: Variation of the adiabatic invariant $J_2$ of the fast oscillator during one ``period'' of the slow oscillator for $\eps=0.1$. This curve is more accurate  to indicate the deviations of the adiabatic approximation.

\vspace*{0.5cm} 
\no FIG.1d: Same as figure 1c but with now $\eps=0.25$. Deviations from adiabatic motions are stronger.\

\vspace*{0.5cm}
\no FIG.2: Classical phase diagram of the anisotropic Heisenberg spin chain corresponding to the hamiltonian (\ref{hamchain})

\vspace*{0.5cm}
\no FIG.3: The $3$-generation Husimi cactus.

\vspace*{0.5cm}
\no FIG.4: The $4$-generation Husimi cactus

\vspace*{0.5cm}
\no FIG.5: The energy barrier separating two discrete  ground-states in the $4$-generation cactus. Spins below site $2$ ({\it i.e.} at sites $4,5,8,9,10,11$) are rotated around $\vec{n}_2$ by an angle $\theta$, giving a smooth path on the ground-state manifold. Note that $\theta=0$ and $\theta=\pi$ correspond to coplanar ground-states. The height of this barrier is around $1\%$ of the low energy excitation which  suggests a quite flat low energy landscape in the  spectrum.

\vspace*{0.5cm}
\no FIG.6: The so called Delta chain with $5$ triangles.

\vspace*{0.5cm}
\no FIG.7: Schematic representation of the action of the linearized Hamiltonian flow and the flow associated to internal rotations.

\eject

\hspace*{3cm}
\psfig{figure=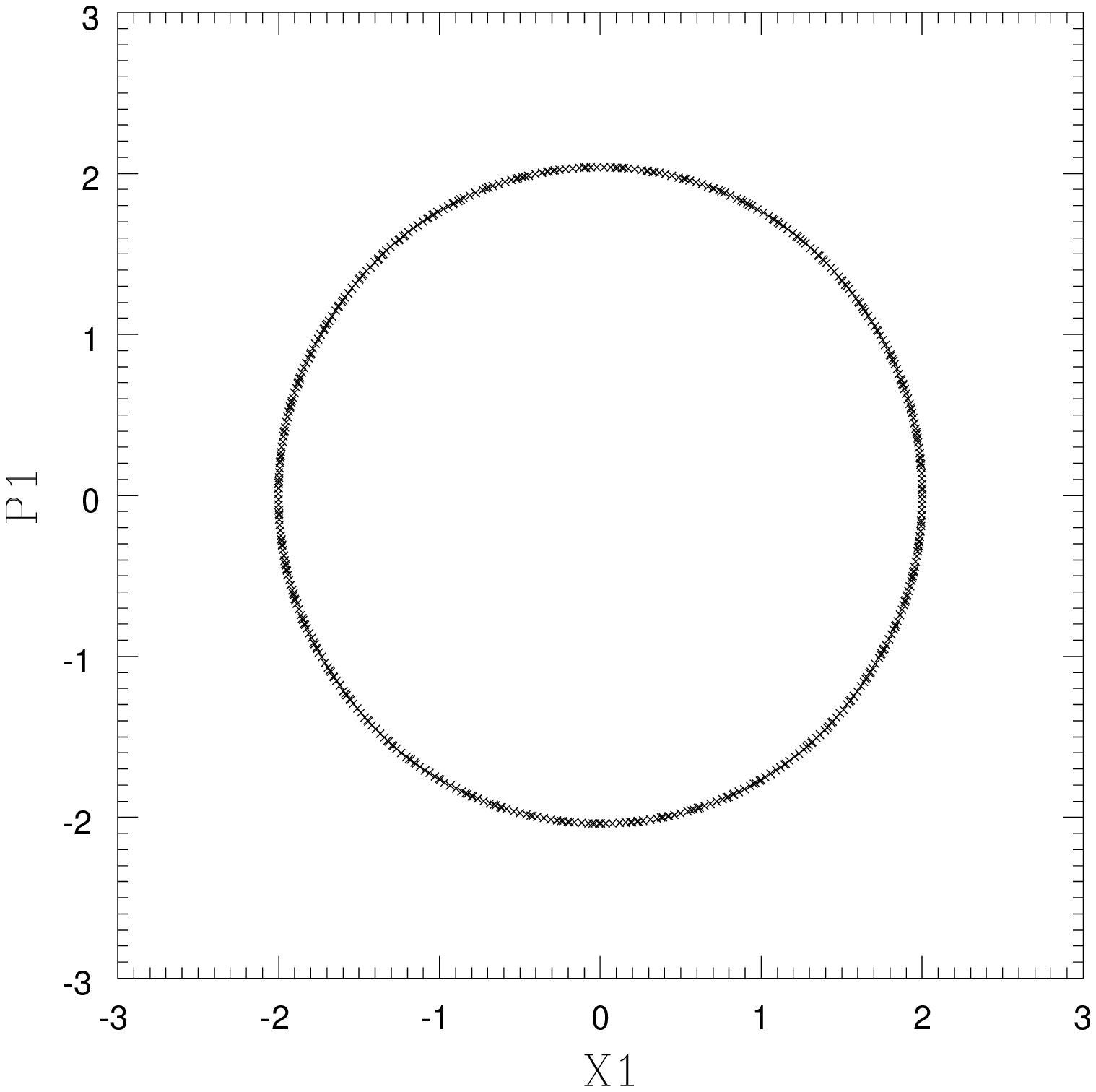,height=10cm,width=10cm}

\vspace*{0.5cm}

\hspace*{3cm}
\psfig{figure=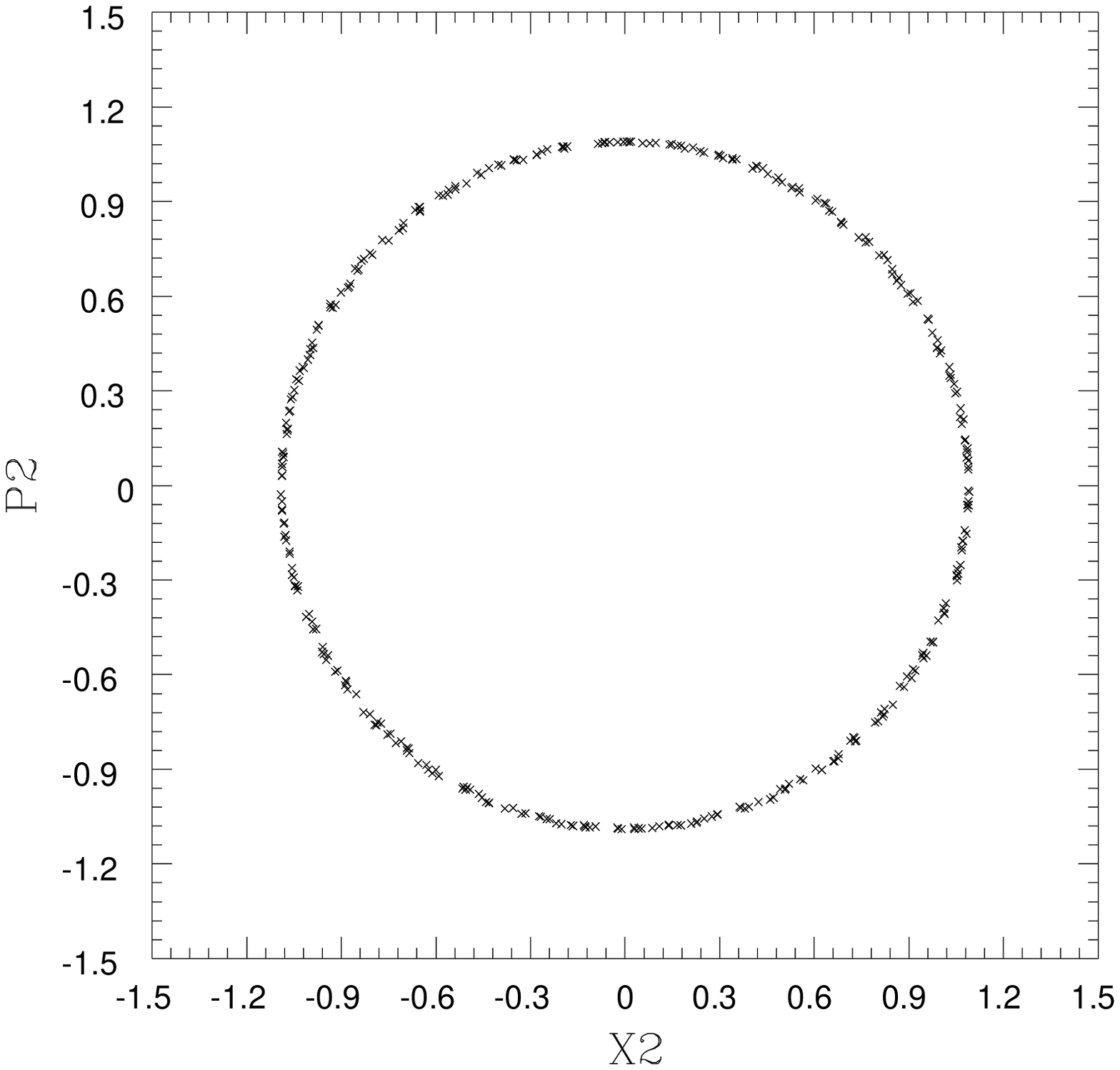,height=10cm,width=10cm}
{\bf Figure 1a} 
\eject

\hspace*{3cm}
\psfig{figure=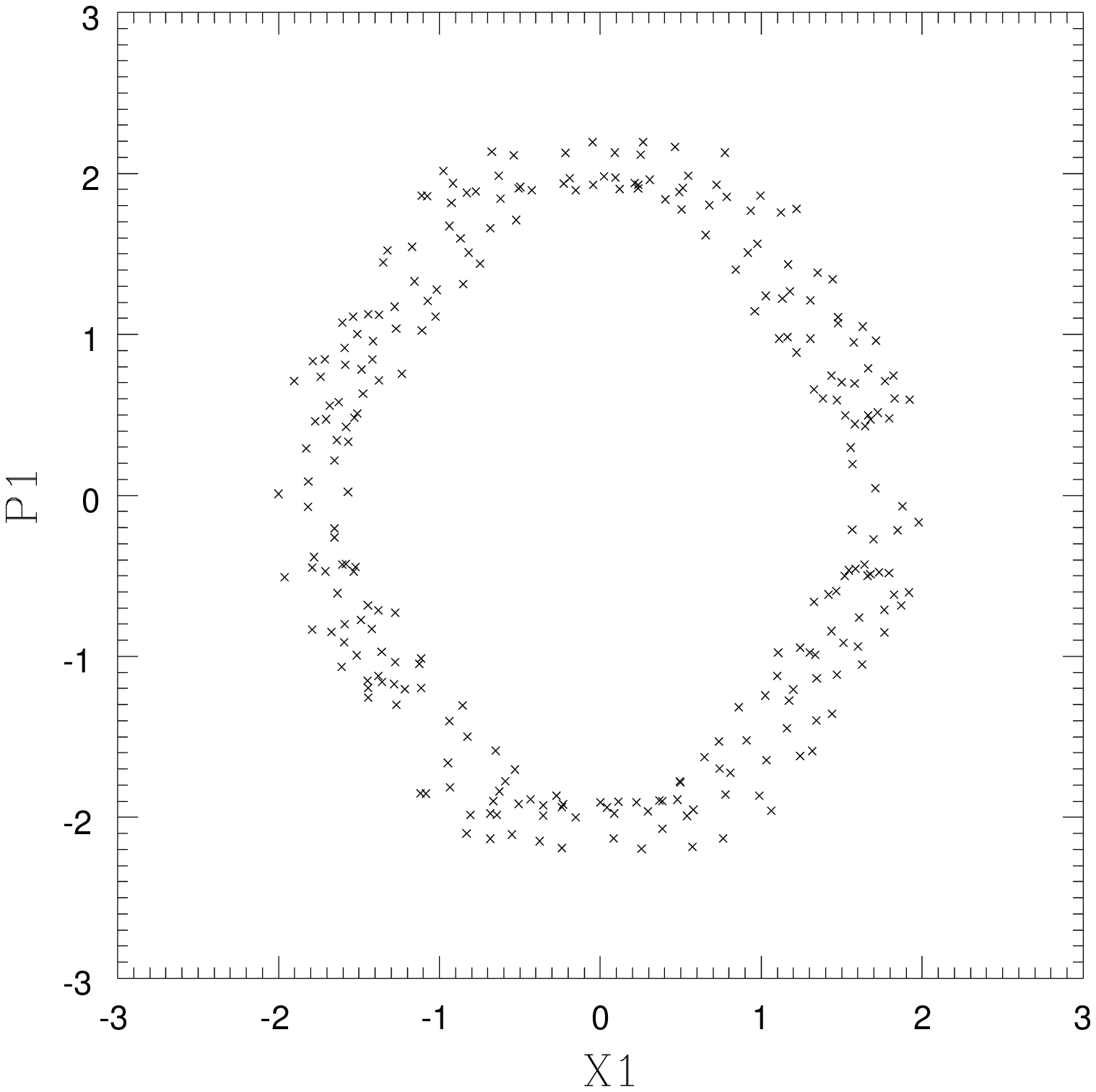,height=10cm,width=10cm}

\vspace*{1cm}

\hspace*{3cm}
\psfig{figure=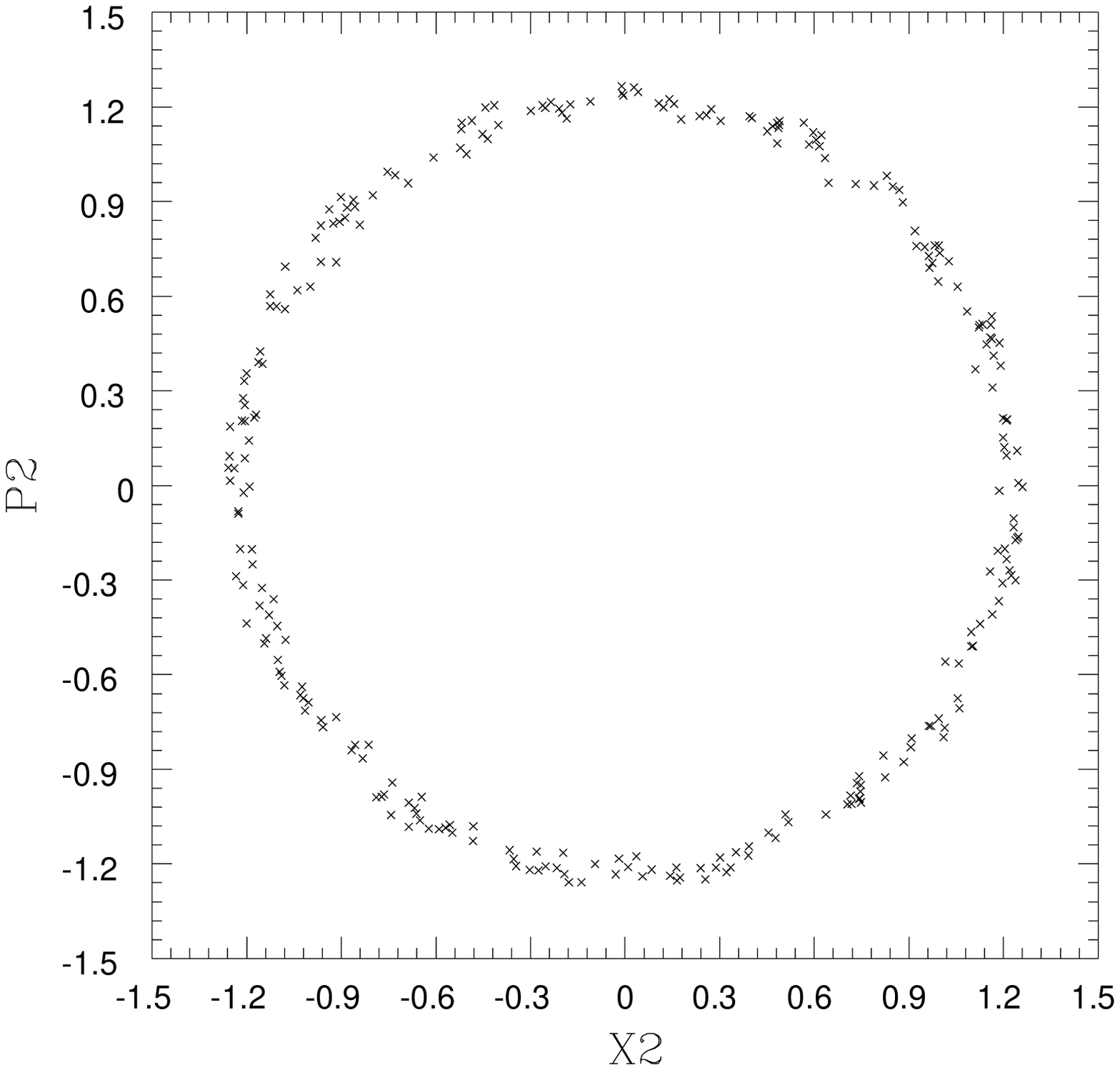,height=10cm,width=10cm}
{\bf Figure 1b} 
\eject

\hspace*{3cm}
\psfig{figure=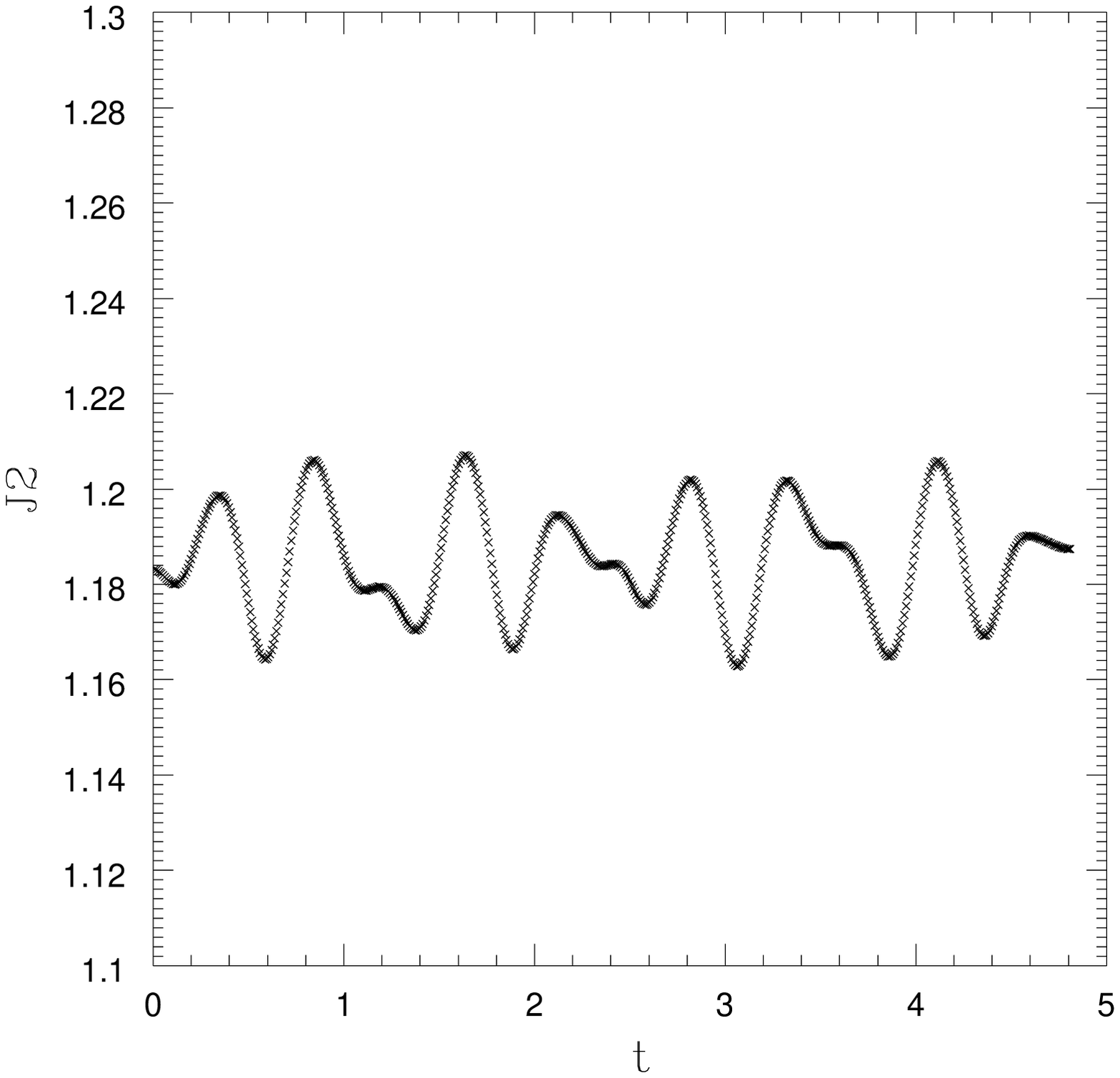,height=10cm,width=10cm}
{\bf Figure 1c} 

\vspace*{1cm}
\hspace*{3cm}
\psfig{figure=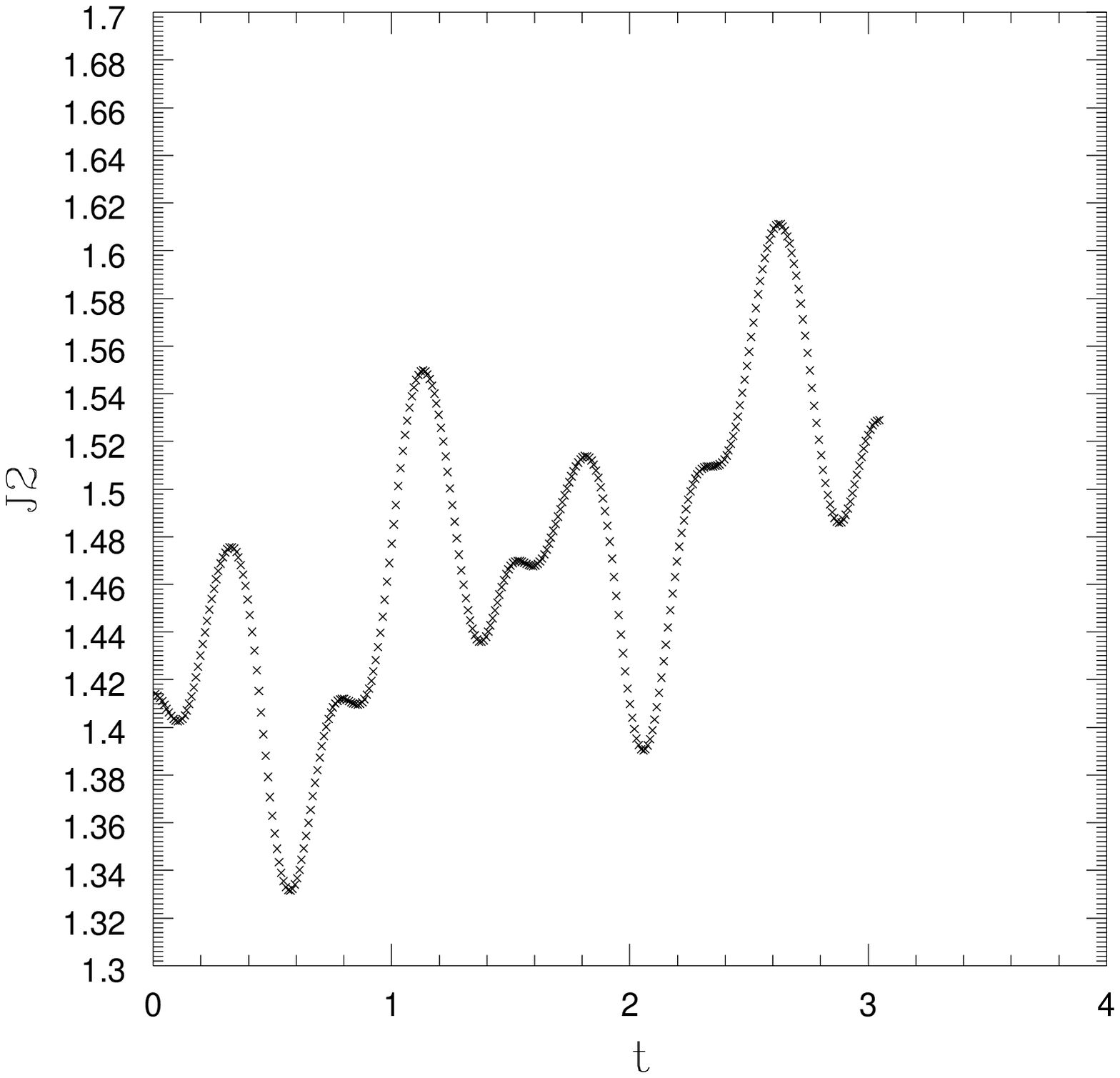,height=10cm,width=10cm}
{\bf Figure 1d} 

\eject

\hspace*{3cm}
\psfig{figure=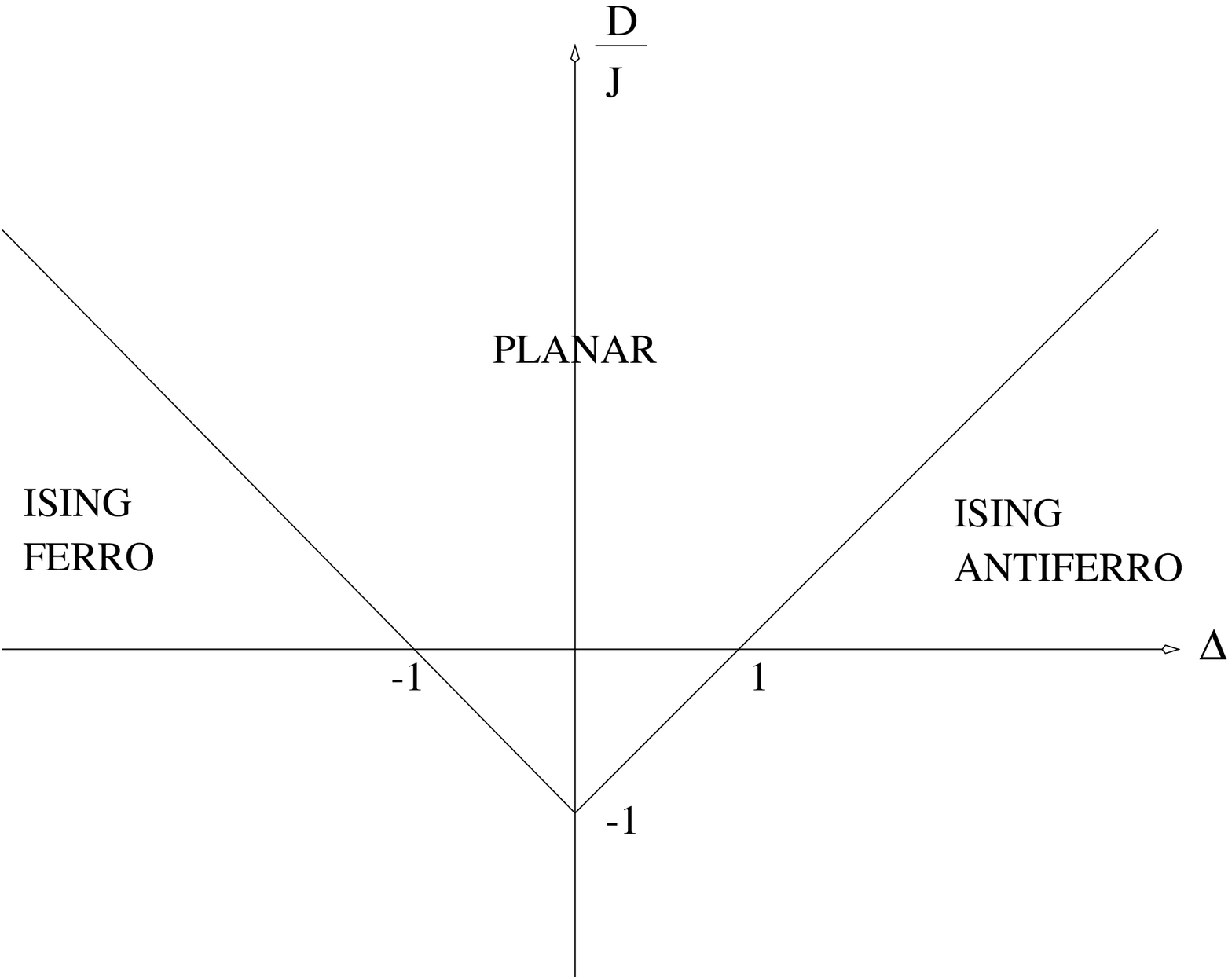,height=10cm,width=10cm}
{\bf Figure 2}

\vspace*{2cm}
\hspace*{3cm}
\psfig{figure=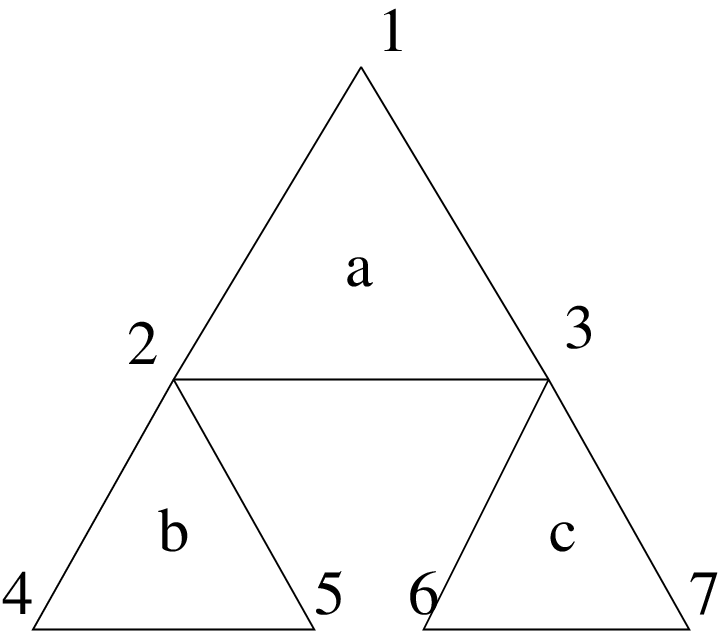,height=10cm,width=10cm}
{\bf Figure 3}

\eject

\hspace*{3cm}
\psfig{figure=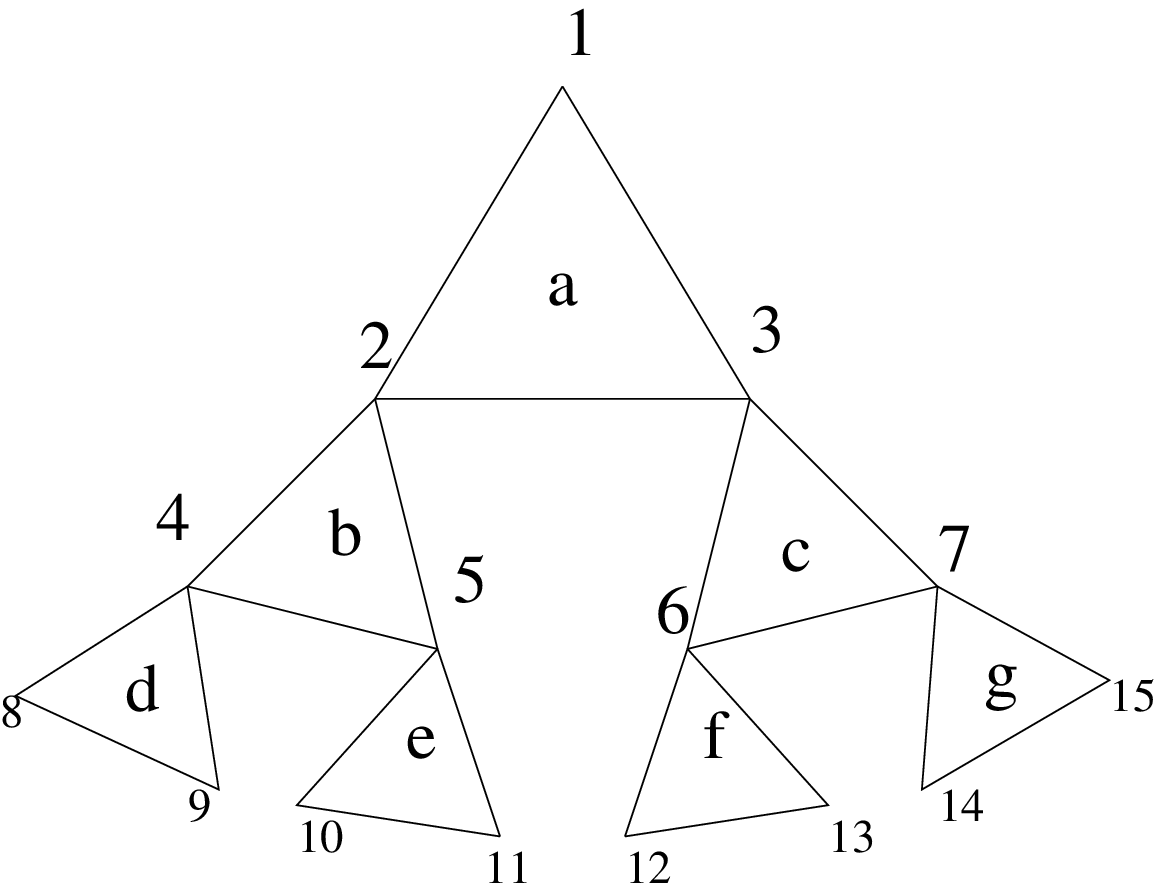,height=10cm,width=10cm}
{\bf Figure 4}

\vspace*{2cm}
\hspace*{4cm}
\psfig{figure=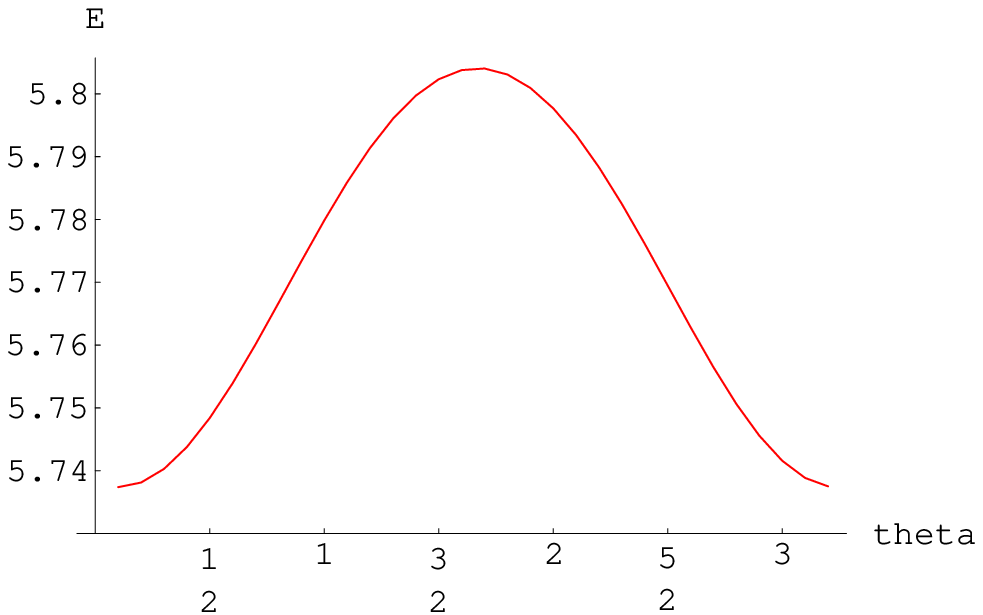,height=10cm,width=10cm}
{\bf Figure 5}
\eject

\hspace*{3cm}
\psfig{figure=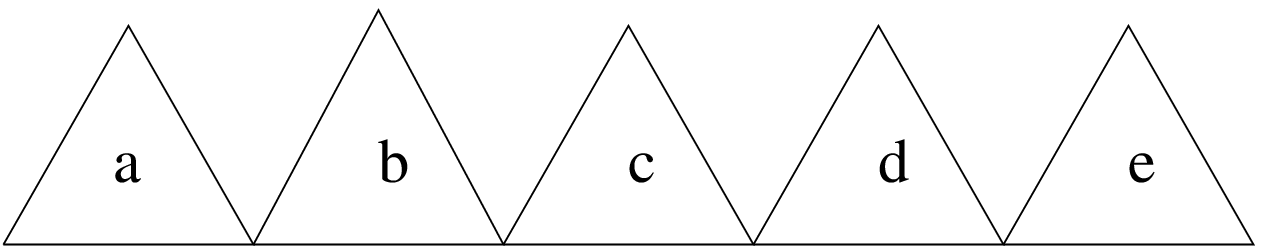,height=3cm,width=10cm}
{\bf Figure 6}

\vspace*{4cm}

\hspace*{3cm}
\psfig{figure=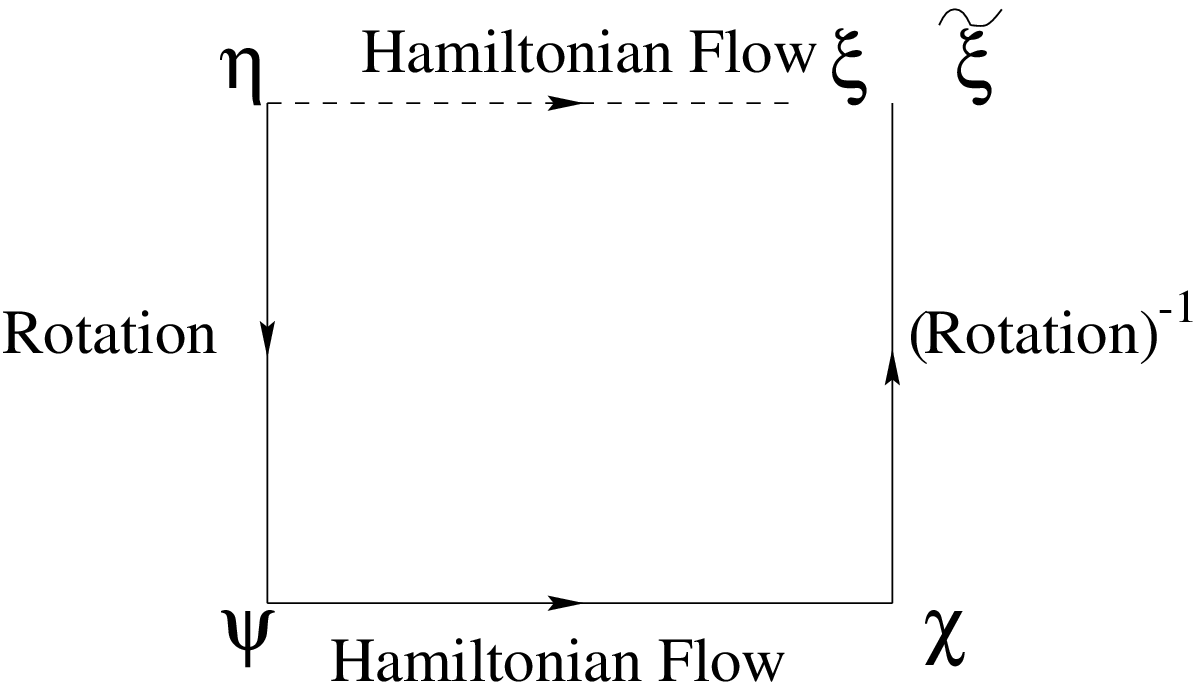,height=8cm,width=8cm}
{\bf Figure 7}

\end{document}